\documentclass[11pt,twocolumn]{article}

\usepackage{paralist}
\usepackage{graphicx}
\usepackage{times}
\usepackage{picinpar}
\usepackage{setspace}
\usepackage{notescnd}
\usepackage{geometry}
\usepackage{fancyhdr}
\newcommand{\ProportionOfText}{0.7}
\renewcommand{\ProportionOfText}{0.7}

\newcounter{remarksCounter}
\begin{document}

\title{{\bf \LARGE The Spatial-Perceptual Design Space: a new comprehension for Data Visualization}\\
{\ }\\
\small{http://dx.doi.org/10.1057/palgrave.ivs.9500161}\\
{\ }\\
{Jose F Rodrigues Jr, Agma J M Traina, Maria C F Oliveira and Caetano Traina Jr}\\
{\it Departamento de Computa\c{c}\~ao}\\
{\it Universidade de S\~ao Paulo}\\
Av. Trabalhador S\~ao-carlense, 400\\ CEP 13560-970 - S\~ao Carlos, SP - Brazil\\
{\bf E-mail:} \{junio,agma,cristina,caetano\}@icmc.usp.br\\
}

\maketitle

\subsection*{\centering Abstract}
{\em \noindent{We revisit the design space of visualizations aiming at identifying and relating its components. In this sense, we establish a model to examine the process through which visualizations become expressive for users. This model has leaded us to a taxonomy oriented to the human visual perception, a conceptualization that provides natural criteria in order to delineate a novel understanding for the visualization design space. The new organization of concepts that we introduce is our main contribution: a grammar for the visualization design based on the review of former works and of classical and state-of-the-art techniques. Like so, the paper is presented as a survey whose structure introduces a new conceptualization for the space of techniques concerning visual analysis.}}\\

\noindent{{\bf keywords: }{Information Visualization, Taxonomy, Design Space, Visualization Model}}\\


\section{Introduction and Related Work}
\label{Introduction}

\noindent{Research on data visualization aims at providing improved mechanisms for information exploration and analysis, offering faster
and friendlier -- if compared to traditional analytical approaches -- techniques to assist on data comprehension. This is possible because an analyst can, with reduced effort, improve his/her understandings of a dataset when interacting through graphical representations.}

Visualization, as it occurs with any research field, will benefit from a conceptual framework to organize methods and techniques in a unified comprehension space, i.e., a delimited space of possibilities where one can navigate through the constituent variables without loosing her/his referential locus. Following this assumption, recent highly relevant publications about the next steps for visualization research have addressed the importance of the topics dealt in the present work. Thomas and Cook \cite{49}, recommend the need to ``Conduct research to formally define the design spaces that capture different classes of visualizations''; Johnson {\it et al.} \cite{JohnsonChallenges06} state, among their long term recommendations, the need to systematically explore the design space of visual representations; and Chaomei \cite{ChaomeiTop05} observes the need to understand elementary perceptual-cognitive tasks based on a substantial amount of empirical evidence.

Several works in the literature have sought to accomplish such task, either from an analytical or from a taxonomical perspective. The analytical approaches listed in Table \ref{table:Works}(a) strive to empirically identify the ultimate elements of visualizations and to draw their relationship. Bertin \cite{44} presents eight visual variables: the two planar dimensions $x$ and $y$ plus six retinal variables: size, value, grain, color, orientation, and shape. Cleveland and McGill \cite{47} and Mackinlay \cite{46} define, in order to support automatic design, empirical studies over simple data visualization aiming at stating the usefulness of specific visual patterns. Card, Mackinlay and Shneiderman \cite{45} follow the theories of Bertin introducing concepts about the importance of the spatial substrate and how it can be used.

\hyphenpenalty=10000
\begin{table*}[htb]
  \begin{center}
  \caption{previous work on conceptual frameworks for visualization.}
  \label{table:Works}
  \end{center}
  \begin{tabular}{|p{6.18in}|}
    \hline
          {\footnotesize{Part(a) - Analytical approaches}}\\
    \hline
  \end{tabular}

  \begin{tabular}{|p{1.01in}|p{4.99in}|}
    \hline
          {\footnotesize{\it{Author}}}
         &{\footnotesize{\it{Contribution}}}\\
    \hline
    \hline
  \end{tabular}

\begin{tabular}{|p{1.0in}|p{4.98in}|}
      {\footnotesize{Bertin \cite{44}}}
     &{\footnotesize{Pioneer in stating that there is a limited set of components that define visual structures: marks and the properties these marks .}}\\
    \hline
      {\footnotesize{Cleveland and McGill \cite{47}}}
     &{\footnotesize{Extends the work by Bertin defining a method for automatic design of presentations based on the evaluation of graphical patterns.}}\\
    \hline
      {\footnotesize{Mackinlay \cite{46}}}
     &{\footnotesize{Following Cleveland and McGill, Mackinlay evaluates graphical patterns to determine the layout directives for data visualization. Adds connection ``-'' and enclosure ``[]'' concepts.}}\\
    \hline
      {\footnotesize{Card, Mackinlay and Shneiderman \cite{45}}}
     &{\footnotesize{Compiles previous works to introduce a general design space oriented to the visual patterns and to the space substrate.}}\\
    \hline
    \hline
\end{tabular}

  \begin{tabular}{|p{6.18in}|}
    \hline
          {\footnotesize{Part(b) - Taxonomical approaches}}\\
    \hline
  \end{tabular}

  \begin{tabular}{|p{1.01in}|p{4.99in}|}
    \hline
          {\footnotesize{\it{Taxonomy}}}
         &{\footnotesize{\it{Criteria}}}\\
    \hline
    \hline
  \end{tabular}

  \begin{tabular}{|p{1.0in}|p{1.5in}|p{1.55in}|p{1.6in}|}
      {\footnotesize{By Keim \cite{14} }}
     &{\footnotesize{1D, 2D, multiD, text/web, hierarchies/graphs and algorithm/software }} 
     &{\footnotesize{standard 2D/3D, geometrical, iconic, dense pixel and stacked }}
     &{\footnotesize{standard, projection, filtering, zoom, distortion and link\&brush }}\\
    \hline
\end{tabular}

  \begin{tabular}{|p{1.0in}|p{2.4in}|p{2.415in}|}
      {\footnotesize{Task by Data Type by Shneiderman \cite{29}}}
     &{\footnotesize{One, two, three, multi-dimensional, tree and network}}
     &{\footnotesize{overview, zoom, filter, details-on-demand, relate, history and extract}}\\
    \hline
\end{tabular}

\begin{tabular}{|p{1.0in}|p{2.4in}|p{2.415in}|}
      {\footnotesize{Data State Reference Model by Chi \cite{ChiTaxonomy00}}}
     &{\footnotesize{Stages of value, analytical abstraction, visualization abstraction and view}}
     &{\footnotesize{transformation operators of data, visualization and visual mapping}}\\
    \hline
\end{tabular}

  \begin{tabular}{|p{1.0in}|p{4.98in}|}
      {\footnotesize{Lohse {\it et al.} \cite{48}}}
     &{\footnotesize{Rectilinear cartesian coordinate graphs, bar graphs, line graphs, matrix diagrams, trilinear charts, response surfaces, topographic charts and conversion scales}}\\
    \hline
\end{tabular}
 \begin{tabular}{|p{1.0in}|p{2.4in}|p{2.415in}|}
      {\footnotesize{Taxonomy draft by Grinstein \cite{9}}}
     &{\footnotesize{Geometric, symbolic, 2D/3D and static/dynamic}}
     &{\footnotesize{browsing, sampling, indirect, associative and system oriented}}\\
    \hline
\end{tabular}

\begin{tabular}{|p{1.0in}|p{4.98in}|}
      {\footnotesize{Digital Visualization Space by Bugajska \cite{BugajskaThesis03}\cite{BugajskaFramework05}}}
     &{\footnotesize{Object (a graphic representing concepts), context (a graphic space representing relations between elements), order (spatial arrangement in the graphic space), goals (structure, organization, appearance) and tasks (overview, zoom, filter, detail-on-demand, relate, history, and extract)}}\\
    \hline
\end{tabular}

\begin{tabular}{|p{1.0in}|p{4.98in}|}
      {\footnotesize{Externalizations by Tweedie \cite{TweedieExternalizations97}}}
     &{\footnotesize{Purpose (display structures, view relations, database queries, ...), data types (values, metadata, structures, derived values), representation (spreadsheets, graphs, retinal variables, structures, multiple views, multi scale and distorted views, ...), interactivity (direct/indirect) and I/O representation}}\\
    \hline
\end{tabular}

\end{table*}
 \hyphenpenalty=50

Taxonomical approaches, listed in Table \ref{table:Works}(b), aim at empirically identifying common characteristics pertaining to existing visualization techniques, in order to propose class-oriented organizations. Keim \cite{14} concentrates on high-level visual patterns in order to define a space of possibilities intuitive for users. Shneiderman \cite{29} focuses on the possible combinations of visualization practices and interaction. Chi \cite{ChiTaxonomy00} describes visualization techniques focusing on data and its transformations. Meanwhile, we focus on organizing the visual features of design from a cognitive perspective. Bugajska \cite{BugajskaThesis03} thoroughly treats the topic of spatial design for abstract visualization proposing a holistic approach in sharing expertise among visual design, computer science, and social fields of study. Tweedie \cite{TweedieExternalizations97} suggests a number of recommendations for design in a work oriented to guidelines.

	Former analytical and taxonomical approaches focus mainly on high-level components of visualizations: data types, visual patterns, visual appearance, user tasks and interaction mechanisms. These works brought significant contribution to visualization design and understanding, but they have overlooked the process through which visualizations become expressive for users. This process relates to how data translates into visual stimuli and to how these stimuli translates into reasoning. 
	Our claim in the present work is that the consideration of this particular process contributes to a better understanding of visual data presentations. With this consideration in mind, we organize the space of visualizations both taxonomically and analytically (design space definition) in a scenario driven by the process of visually expressing information. 

We observe that, distinctly from works that focus on {\it design guidelines}, analytical works focusing on design space theories aim at providing a better organization of concepts so that design-related activities can be better engendered. In this context, our work defines a {\it design space} in order to introduce a new conceptual structure, a grammar, related to the domain of design possibilities. On the other hand, regarding the important topic of design guidelines, MacEachren \cite{MacEachrenMaps04} presents a complete book that extensively reviews methods for data presentation at the same time that he evaluates their adequacy and utilization. His work, a rich review of practices for data visual presentation, is a cornerstone reference for visualization design. MacEachren states that the meaning of a map is not absolute but a product of the society and its culture. He also states that the way information is mentally represented determines how groups and societies can develop a consensus about data representation. This consensus, in the realm of data visualization, is the object of the theoretical and empirical investigation that supports our work.

In the sphere of design space theories, our initial considerations are based on the work by Card {\it et al.} \cite{45}, who identify the elements of visualizations. Card {\it et al.} starts from the assumption that visualizations are limited to the spatial substrate, to marks and to graphical properties pertaining to these marks. Such graphical properties are what Bertin [3] defines as retinal properties. Ware [39] distinguishes the retinal properties proposed by Bertin as visual patterns defined either by shape or by color.
	In line with these former studies, we consider visualizations as being composed of three fundamental components: position, shape and color. Each of these components conveys visual perceptual cues, which we refer to as {\it visual perceptions}.
	The observation of these phenomena is formalized in a model named {\it Visual Expression Process}, which grounds this work.

	The remaining of this paper is organized as follows. 
	Section \ref{sec:Fundamentals} introduces the fundamentals of this work, presents the {\it Visual Expression Process}, and delineates our research line. 
	Section \ref{sec:Our_Taxonomy_Proposal} guides our initial ideas by proposing the {\it Spatial-Perceptual Taxonomy}. 
	Section \ref{sec:Analytical_examples} exemplifies our ideas and discusses further possibilities, while section \ref{sec:DesignSpace} delineates the {\it Spatial-Perceptual Design Space}. 
	Section \ref{sec:InteractionTechniques} explains the role of interaction in the proposed framework. 
	Section \ref{sec:GeneralVisualization} proposes a unified model for visualization, named {\it Visualization Machine}, which integrates all the proposed concepts. 
	Section \ref{sec:Conclusions} presents the concluding remarks.

\section{Fundamentals}
\label{sec:Fundamentals}

	\noindent In the present work we define two levels of abstraction for the practice of visualization: design of visualization techniques and design of visualization systems. While the design of techniques is concerned to how a visual representation should be structured and look like, the design of visualization systems is also concerned to the aspects of data processing and interaction.
	We emphasize that {\em in the discussion that follows, the scope of this work applies essentially to the topic of visualization techniques. Nevertheless, for the sake of completeness, we touch the topic of visualization systems (processing + visualization + interaction) discussing how and where our theory relates to systematization}. In this section, initially, we introduce the underlying knowledge and the basic definitions central to the development of our ideas.

\subsection{Semiotics and Pre-attentive Visual Stimuli}
\label{subsec:PreAttentiveVisualStimuli}

\begin{figure*}[htb]
    \centering
\includegraphics[width=\ProportionOfText\textwidth]{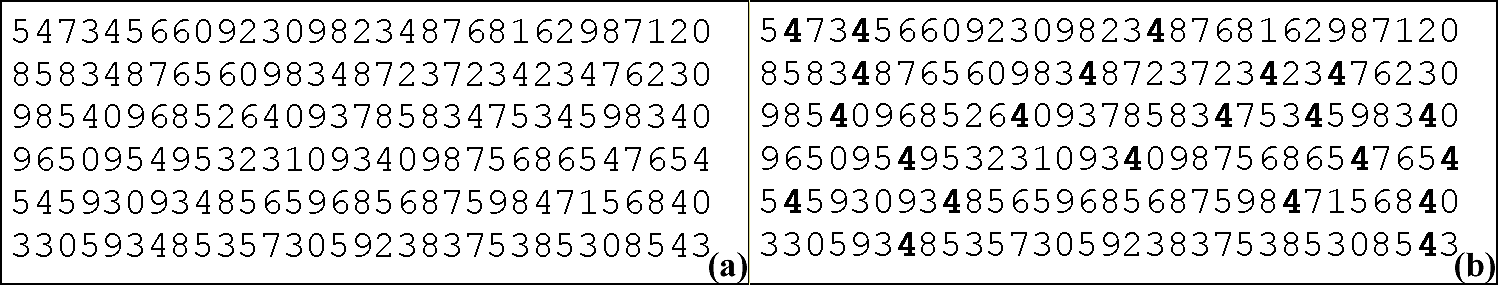}
    \caption{Example of pre-attention. Identifying the number of ``4'' symbols (22 in total) is easier when the symbols are emphasized (b), than when they are not (a). Adapted from Ware \cite{35}.}
    \label{fig:PreAttentive}
\end{figure*}

\noindent{Semiotic Theory, from Vision Science, is the study of signs and their ability to convey meaning. According to this theory, the visual process comprises two phases, namely, the parallel extraction of low-level properties (called pre-attentive processing), followed by a slower detailed scan. The first phase, pre-attentive processing, plays a crucial role in promoting the major benefit of visualizations, that is, improved and faster data comprehension \cite{43}. Meanwhile, the second phase addresses conventional reading practices that do not contribute towards faster visual cognition. In this sense, Ware states that understanding what is processed pre-attentively, see Figure \ref{fig:PreAttentive}, is probably the most important contribution that Vision Science can bring to data visualization \cite{35}.}

\begin{figure*}[htb]
    \centering
\includegraphics[width=\ProportionOfText\textwidth]{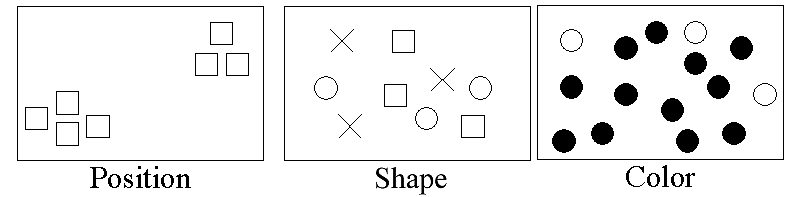}
    \caption{Pre-attentive visual stimuli.}
    \label{Semiotical}
\end{figure*}

	Pre-attentive processing refers to whatever can be visually identified through unconscious processes. As such, it determines which visual objects are instantly and effortlessly brought to our
attention.
	As depicted in figure \ref{Semiotical}, Ware [39] identifies the categories of visual features that are pre-attentively processed. His study considers Position, Shape and Color -- besides the animation of each of these three factors.
	According to Pylyshyn {\it et al.} \cite{23}, there are specialized areas of the brain to process each of these stimuli. Actually, the position-shape-color perception is true for everything we see. At any moment and for anything on which we focus our eyes, we can ask three questions: where is it? what is its shape? and what color is it?

For visualizations to be effective, they must build on pre-attentive features in order to maximize the number of just noticeable differences. In fact, although visual processes are not limited to pre-attention, visualization design is supposedly oriented to maximize design-specific pre-attentive effects. That is, there are many graphical properties, but just a limited number of them can be used for visual analysis \cite{45}. Accordingly, in this research we assume that visualizations are composed of features that are potentially or desirably pre-attentive.

Following this conception, pre-attention effectiveness may vary, or even be absent, depending on the data and on the particular design.\\

\subsection{Spatialization and Position}
\label{subsec:Spatialization_and_position}

\noindent{According to Gattis \cite{GattisThought01}, for a visualization to be effective, organization plays an important role. In order to achieve such organization, the spatial schemas (spatializations) are the main graphic representation. Data spatialization refers to its transformation into a visible/spatial format. Straightly, it is possible to conclude that the spatial positioning, dictated by the spatialization process, is what promotes positional pre-attention. In their state-of-the-art work, Card {\it et al.} \cite{45} refer to the spatial substrate as the most fundamental aspect of a visual structure, being the first decision in the visualization design. In fact, Rohrer {\it et al.} \cite{27} state that visualizing the non-visual requires mapping the abstract into a physical form. Rhyne {\it et al.} \cite{25} distinguishes Scientific visualization and Information visualization based on whether the spatialization mechanism is given or chosen, respectively. Spatialization is the fundamental element to enable the visual data analysis, at the same time that it dictates the characteristics of the pre-attentive positional perception. Indeed, positional expressiveness depends on a mapping function that drives data elements into spatial positions.}

\subsection{The Visual Expression Process}
\label{subsec:VisualExpressionProcess}

\begin{figure*}[htb]
    \centering
\includegraphics[width=\textwidth]{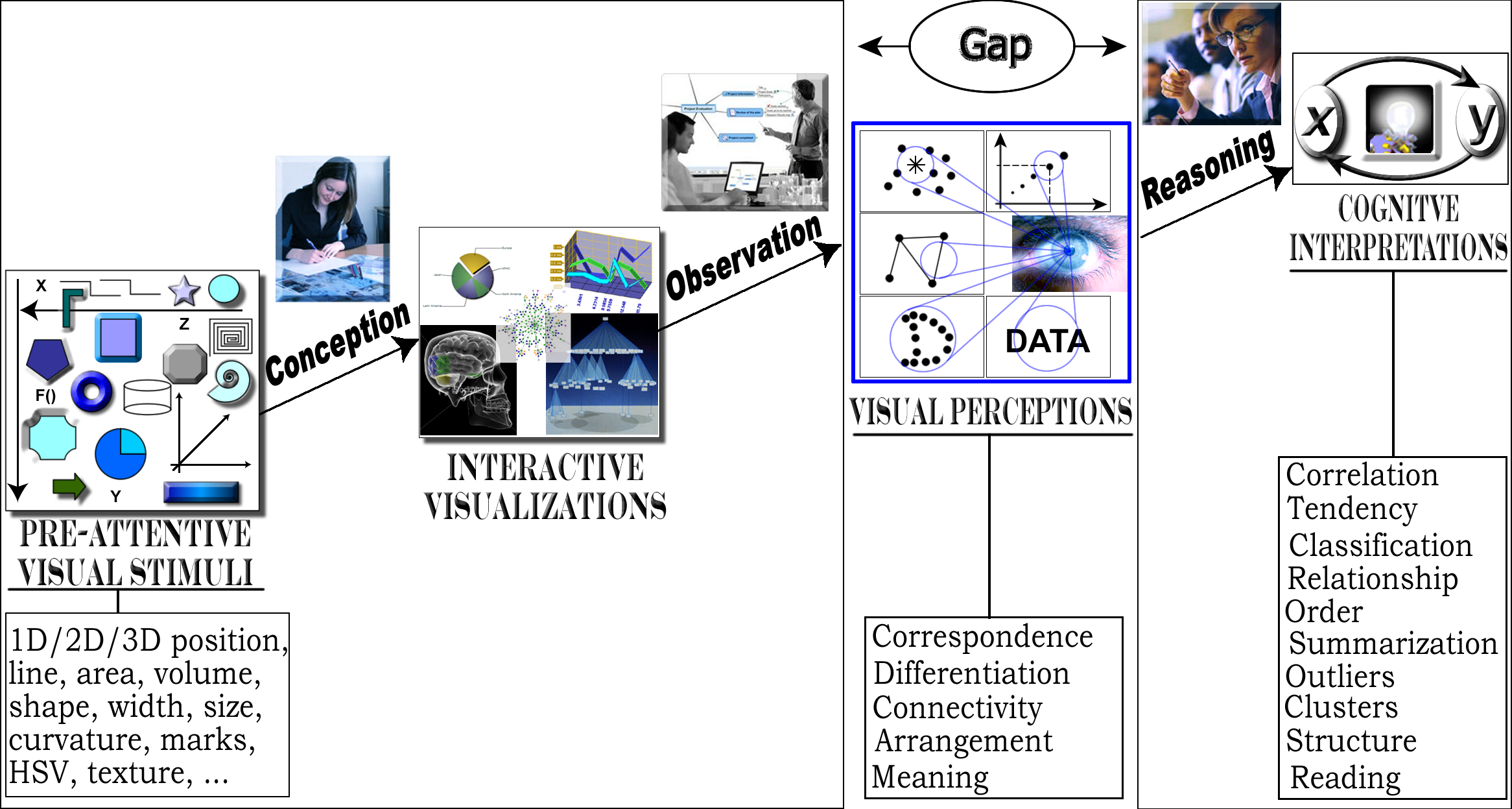}
    \caption{The visual expression process.}
    \label{fig:VisualExpressionProcess}
\end{figure*}

\noindent The {\it Visual Expression Process} is a scheme of how visualizations improve useful knowledge acquisition. The diagram in Figure \ref{fig:VisualExpressionProcess} describes a process that departs from pre-attentive visual stimuli -- the presence of a dataset is naturally assumed -- and culminates in data interpretation. The key elements of such process are: the pre-attentive stimuli used to conceive visualizations, the concept of {\it visual perception}, and the interpretation factor. 
The model is now examined in order to grasp its structure, in search for a new strategy for visualization science.\\

\noindent{\bf Pre-attentive stimuli and visual patterns}

\noindent Our initial assumption is that a visualization is composed of the three pre-attentive stimuli -- position, shape and color -- presented in section \ref{subsec:PreAttentiveVisualStimuli} and depicted at the left-hand side of Figure \ref{fig:VisualExpressionProcess}. 
For each of them, a set of visual patterns may be employed for the visualization design. 
Such patterns include (but are not limited to):\\
\begin{compactitem}
\item{position: 1D/2D/3D position, stereoscopic depth;}
\item{shape: line, area, volume, form, orientation, length, width, collinearity, size, curvature, marks, numerosity, convex/concave;}
\item{color: hue, saturation, brightness.\\}
\end{compactitem}

\noindent{\bf Visual perceptions}

\noindent{\it Visual perceptions} designate the limited number of user-related phenomena that are fired by the pre-attentive stimuli composing a visual presentation. Visual perceptions are understood as the recognition of visual stimuli chiefly based on the memory of the user who observes the scene. They are observable even if she/he has no knowledge about the underlying data and, therefore, they are an inherent component of the visualization practice.

Visual perceptions seem to be the natural features that any user tries to identify when interpreting a visualization scene. If they are not found by the user, the pipeline outlined in Figure \ref{fig:VisualExpressionProcess} is broken and the visual expression process is interrupted. Due to their importance, we have extensively surveyed the occurrence of visual perceptions in visualization literature. We found a limited set and, by empirical observation, we verified that this set is recurrent in visualization techniques. 
We also observed that each visual perception is fired by one or multiple pre-attentive stimuli, having either a discrete or continuous nature.

In the work by Bertin \cite{44}, and also observed by Bowman \cite{BowmanGCommunicatoin68} and Card \cite{45}, one can identify two occurrences of visual perceptions: extended expression (which we call correspondence) and differentiation of marks. Mackinlay \cite{46} states that the perception of relationships comes from visual patterns that remind of the notion of connectivity. Besides these three visual perceptions, namely correspondence, differentiation and connectivity, we also identified: arrangement, for perceptions that arise from group positional configurations (Gestalt principles \cite{KoffkaGestalt35}); and meaning, for perceptions that draw on resemblance with previous knowledge and/or expertise. The list of visual perceptions we have identified includes, but it is not limited to:

\begin{compactitem}
\item{{\it correspondence}: Each position/shape/color establishes a distinct correspondence with respect to some referential map. The need for an explicit visual map or for an implicit mental map is assumed. Examples of visual maps include axes, geographical maps, shape/color dictionaries and position/shape/color variation ranges. 
Examples of mental maps include known orderings and shape metaphors;}

\item{{\it differentiation}: Each position/shape/color discriminates one or more graphical items. Differentiation is the simplest visual perception and can be understood as a kind of correspondence in which the user creates a temporary mental map in memory;}

\item{{\it connectivity}: Shapes that convey information about relationships;}

\item{{\it arrangement}: Gestalt principles of organization [Koffka, 1935]. Positional placements (similarity, continuity, closure, proximity, symmetry and figure/background) that convey information about group properties, e.g. clusters and structural cues;}

\item{{\it meaning}: Positions/shapes/colors whose interpretation relies on the expertise of the user or on previous knowledge. Meaning can be understood as a kind of correspondence established from visual entities to concepts retained in the long term memory of the user. Something is perceived as meaningful if its significance extends beyond the context of the visualization. 
In contrast, the components of a data visualization scene become significant due to their own ensemble.\\}
\end{compactitem}

Other visual perceptions could possibly be identified, as for example, textual labels, textures, and enclosure, as proposed by Mackinlay \cite{46}. However, we do not include them in the above list. Textual labels may be considered compositions of shapes expressing perceptions of meaning or differentiation. As for textures, Kimchi \cite{KimchiHierarchicalPercpt98} reviews the literature of Vision Science stating that textures have been interpreted as similarity-based groupings of shapes. For the visualization science, textures are somewhat at the boundary between shape expression (the individuation of the shapes in the texture) and color expression (the grouped shapes). Typically, two perceptions rise: correspondence/differentiation (like color expressiveness) or meaningful resemblance to a known material surface (like shape expressiveness).\\

\noindent{{\bf Interpretations}}

\noindent{Interpretations refer to conclusions, inferences or deductions produced with the aid of the visualization scene in conjunction with the knowledge of the data domain. As depicted in Figure \ref{fig:VisualExpressionProcess}, the user interpretation of a given visualization occurs as a consequence of her/his visual perception. A data visualization may be more or less effective in providing proper interpretations, it depends on the data, on the visualization itself and on the user. Our understanding is that the interpretations include, but are not limited to the following list of concepts: correlation, tendency, classification, relationship, summarization, outlier, cluster, structure and reading.}\\

\noindent{\bf The visual expression process\\} Figure \ref{fig:VisualExpressionProcess} aggregates the observations made so far into a single process, named {\it Visual Expression Process}. Three cognitive tasks are involved in this process: conception, observation and reasoning. Designing a visualization requires establishing how position, shape and color will be employed for visual stimulation; during observation, visualizations (properly designed) necessarily provide visual perceptions; user reasoning, based on the visual perceptions, may or may not induce interpretations about the data.

Our view is that the Visual Expression Process is the underlying process behind all visualization techniques. Intuitively, one can say that the visual patterns of position, shape and color are purely related to the raw data and that, through visual perception, interpretations can be stimulated. The interpretations, in turn, are related not to data, but to new information in the context of the application domain. In other words, interpretation leads to knowledge. 

The Visual Expression Process establishes that, once a visualization is conceived, two additional steps are required prior to achieving knowledge generation: observation and reasoning. However, a gap separates visualizations from their interpretation. This gap is filled by the concept of Visual Perception, which relies on the user memory in order to transform observations into procedures related to the {\it science of analytical reasoning} \cite{49}, and to knowledge.  In this work we identify how the Visual Expression Process establishes key issues for visualization research: what are the elements of visualization techniques? how such elements can be used in order to define analytical visual perception? what factor bridges visual stimulation and knowledge production?

The foundations and argumentations we develop in this work apply to empirical observations related to vision processes seeking to maximize the perception for data exploration and, as such, they are restricted to the science of Data Visualization -- we do not claim that they apply to vision science in general.\\

\subsection{Research Line}
\label{subsec:Proposal}

\noindent{The Visual Expression Process  presented in the former section establishes that, given a visualization, two steps are required to produce knowledge: observation of visual stimuli and reasoning. Between these two steps there is a gap separating visualizations from interpretations. This gap is fulfilled with the concept of Visual Perceptions.}

Our thesis is that it is possible to deploy a taxonomical and analytical theory that, while based on visual stimuli, is oriented to visual perceptions. This choice is justified by the observed properties of Visual Perceptions, that:

\begin{itemize}
\item{Are common to every visualization, in contrast to interpretations;}
\item{Are not numerous, in contrast to visual patterns;}
\item{Can be categorically related to the pre-attentive stimuli, allowing for immediate recognition and association;}
\item{Constitute a key element for visual analytical cognition.}
\end{itemize}

These features suggest that a perceptions-oriented-study can be more rational than former works mainly oriented to visual patterns (see Section \ref{Introduction}). That is, it is easier to know which interpretations can be achieved from a given perception than from a given visual pattern. Our point is that it is possible to break a visualization into distinct parts: pre-attentive stimulli of position (spatialization), color and shape that, in turn, derive visual perceptions of correspondence, connectivity, differentiation, arrangement and meaning. We believe that this analytical course can provide a simpler control of which visual effects are effective, and for what. We use this notion in order to review the visualization science from a non-holist perspective.

The goal of this work is to introduce an alternative organization of concepts, a platform for comprehension and for discrete analysis of visualization techniques. In this sense, the paper starts with the description of a taxonomy that identifies the features that characterize visualization techniques. These features are, then, represented in a space of design possibilities where each set of points corresponds to a taxonomical class. Complementing our theorization, we introduce an ideal model for navigating the design possibilities predicted by our theory.\\

\subsection{Contributions}
\label{subsec:contributions}

\noindent{To state the contributions of this work, we compare its principles with the work by Card {\it et al.} \cite{45}, which is the most cited proposition for the topic of visualization design space. Compared to Card {\it et al.} we make the following observations:}

\begin{itemize}
	\item{Distinctly from Card {\it et al.} and from most of prior theories, our work is not centered on visual patterns (retinal properties) whose number and diversity raise difficulties for a clear and general analytical view of a given visualization. Instead, we argue that a user is more interested on the visual perceptions that she/he can benefit of than on the visual patterns that she/he can use. For example, instead of desiring a set of axes and points, a user would be more interested in observing properties of correspondence that are present (or not) in the data;}

\item{Card {\it et al.} affirm that the space can be configured through techniques of composition, alignment, folding, recursion and overloading. These items reveals that, instead of defining a general understanding, the authors tried to exhaustively list the different ways that the space can be occupied in visualization design. This lack of generality prevents the analyses of techniques such as the Star Glyphs and VisImpact \cite{HaoBProcess06}. Differently, we introduce an alternative comprehension for space utilization, which is closely related to spatialization and to gradual occupation;}

\item{In the same way as for spatial design, interaction is not part of Card {\it et al.} theory (nor of many prior theories). Instead, the authors state an exhaustive list of observable interactive practices. In their work there is no relation between the visualization elements and the interaction practices that they identify. In a more general way, our proposition states that interaction mechanisms are natural ways to alter the parameters of the pre-attentive components present in any visualization.}
\end{itemize}

\section{The Spatial-Perceptual Taxonomy}
\label{sec:Our_Taxonomy_Proposal}

\noindent{Classification serves as an instrument for the structured development of new guidelines, as well as to organize the existing ones \cite{BugajskaThesis03}. Like so, the objective of the following taxonomical study is to improve the comprehension of what are the components of visualizations and how to classify such components from a cognitive/perceptual perspective. Rather than focusing on specific visualization cases, the discussion fosters a comprehension for the analytical appreciation of the overall visualization practice, independent of usability or adequacy.}

We start our taxonomy by discussing spatialization. In the next sections we discuss how shape and color are employed by visualization techniques.

\subsection{Spatialization}

\noindent{Based on the notion of spatialization, we have verified that visualization techniques can be grouped according to how they are spatialized. We have identified the following classes for the criterion of spatialization: Structure Exposition, Sequencing, Projection and Reproduction.}

\begin{figure*}[htb]
    \centering
\includegraphics[width=\ProportionOfText\textwidth]{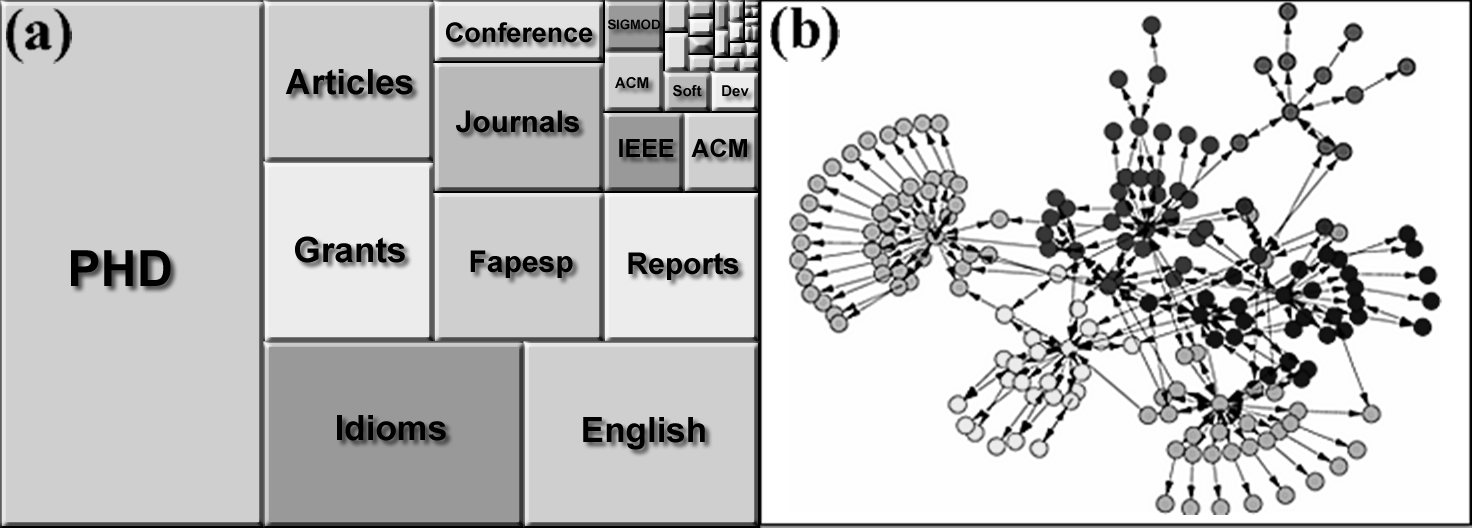}
    \caption{(a) TreeMap structure exposition. {\it Position:} hierarchical arrangement;
    {\it shape:} continuous correspondence (size proportionality); {\it color:} discrete differentiation.
    (b) Force-directed structure exposition. {\it Position:} relational arrangement;
    {\it shape:} connection lines and meaningful arrows; {\it color:} discrete differentiation.}
    \label{StructureExposition}
\end{figure*}

\noindent{$\bullet${\it\ Structure exposition}: data can embed intrinsic structures, such as hierarchies or relationship networks (graph-like), that embody a considerable part of the data meaning. We designate as Structure Exposition the techniques aiming at exposing the data structure. Such techniques rely on methods to adjust the data presentation so that the underlying structure can be visually noticed. They are usually domain specific and attempt to maximize the perception of arrangement. Examples include the TreeMap technique \cite{30} (hierarchical recursive positioning), illustrated in Figure \ref{StructureExposition}(a), and force-directed graph layouts \cite{41} (iterative positioning), illustrated in Figure \ref{StructureExposition}(b);}

\begin{figure*}[htb]
    \centering
\includegraphics[width=\ProportionOfText\textwidth]{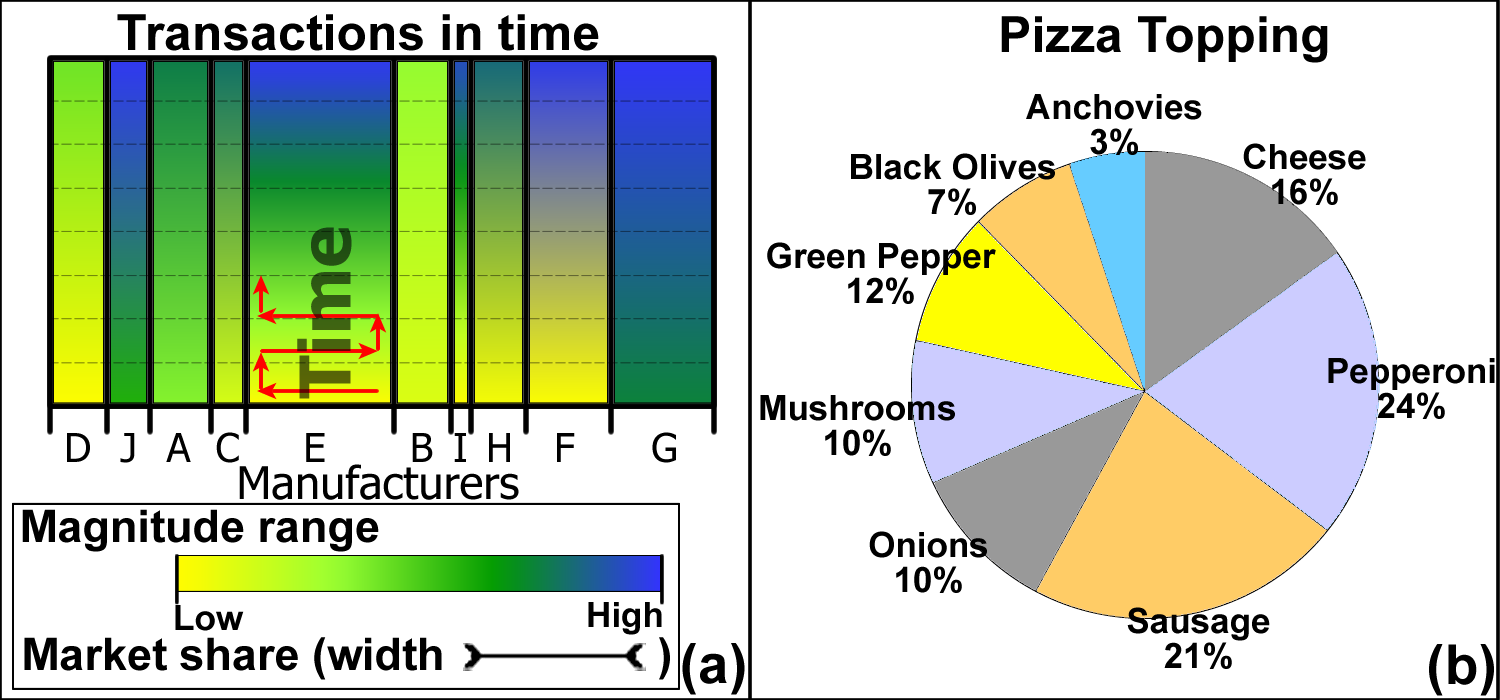}
    \caption{(a) Pixel Bar Charts - each pixel maps sequentially one manufacturerӳ transaction. Position: discrete correspondence to the map of labels that identify the manufacturers positioned in horizontal sequence, and continuous correspondence of the transactions (pixels) to the sequential mental map following the ascending time order declared over the scene; shape: continuous correspondence (width); color: continuous correspondence. (b) Pie chart sequencing. Position: discrete correspondence to the non-ordered circular sequence of labels; shape: continuous correspondence (area); color: discrete differentiation.}
    \label{Sequencing}
\end{figure*}

\noindent{$\bullet${\it\ Sequencing}: is the simplest positioning procedure. It works by placing the dataset in a sequential arrangement, typically following the overall equation:}

\begin{equation}
(x_{i+1},y_{i+1},z_{i+1})=f(x_{i},y_{i},z_{i})
\label{eq:Sequencing}
\end{equation}
\noindent that is, the position occupied by a particular data item only depends on the position of the preceding item. Sequential spatializations adopt linear, circular or more elaborated positioning patterns (arrangements) that constitute the positional mapping of the visualization. These arrangements can fire visual perceptions of differentiation or correspondence. Differentiation is intrinsic to position, while correspondence is achieved when a map is provided. Sequential spatializations typically adopt mental or visual positional maps. Mental mappings are based on a declared order, as illustrated in Figure \ref{Sequencing}(a). Visual mappings are based on a positional map that explicitly identifies each position, as illustrated in Figure \ref{Sequencing}(b).

Sequential positioning techniques tend to fully populate the display area and some of them are referred to as dense pixel
displays. 
The pixel-oriented techniques proposed by Keim \cite{13} are well-known examples of sequential positioning. They use just color and no shape encoding to present the data items, which are positioned according to elaborated sequential patterns \cite{15}. Pixel Bar Charts \cite{16} is a variation of such techniques, it relies on two spatialization cycles in order to benefit from size encoding.             

\begin{figure*}[htb]
    \centering
\includegraphics[width=\ProportionOfText\textwidth]{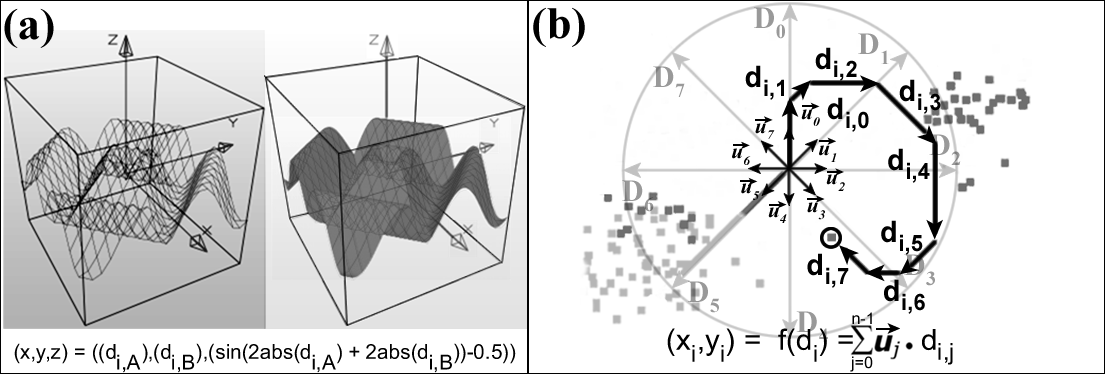}
    \caption{3D functional projection. Position: continuous correspondence to the projection axes; meaningful lines implicit interpolation (left), and surface explicit interpolation (right); shape: not given/chosen; color: none. (b) Star Coordinates 2D projection of 8-dimensional data. Position: continuous correspondence to the projection axes; shape: none; color: discrete correspondence to selections over axis $D_5$.}
    \label{Projection}
\end{figure*}

\noindent{$\bullet${\it\ Projection}: stands for a data display modeled by functional variables. The position of
    a data item is defined by a mathematical function (either explicit or implicit) that generates a set of positional marks representing the data. The marks state perception of correspondence to the axes used as positional maps. At the same time, the discrete set of positional marks can compose, via implicit or explicit interpolation, lines, areas, surfaces or volumes that induce perception meaning. An axial reference is required for spatializations based on projection that, in general, take the format:}

\begin{equation}
(x_i,y_i,z_i)=f(d_{(i,0)}, d_{(i,1)}, ..., d_{(i,n-1)})
\label{eq:Projection}
\end{equation}

\noindent where $d_{i,j}$ is the $j-th$ attribute of the $i-th$ data item and
$n$ is the dataset dimensionality. Examples include Parallel Coordinates
(one projection per data dimension), conventional plots and Star Coordinates \cite{12}, as illustrated in Figures \ref{Projection}(a)
and \ref{Projection}(b);

There is a well-defined boundary between spatializations based on projection and those based on sequencing.
Projections are sequences strictly limited to quantitative data and that have been restricted to positional maps in the format of axes. Sequencings apply primarily to nominal and ordinal data and may have arbitrary positional maps. Hybrid variations exist that benefit from both spatializations simultaneously, as discussed further.             

\begin{figure*}[htb]
    \centering
\includegraphics[width=\ProportionOfText\textwidth]{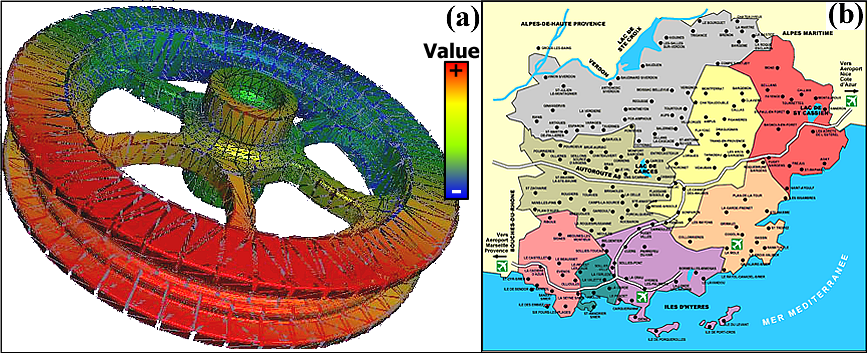}
    \caption{(a) Rendered dataset reproduction. Position: the set of reproduced positional marks compose a meaningful volume via explicit interpolation of the points; shape: not given/chosen; color: continuous correspondence. (b) Geographical map reproduction. Position: correspondence to the background map used for reproduction (the data positioning is known beforehand); shape: differentiation (large dots), and meaningful airport and road identifiers; color: discrete differentiation, and meaning following the knowledge that water is blue. (a) reproduced with permission granted by S.G. Eick.}
    \label{Reproduction}
\end{figure*}

\noindent{$\bullet${\it\ Reproduction}: data positioning is known beforehand, having been determined by the original spatialization of the system/phenomenon that generated the data. Ideally, the visualization and the observed phenomenon should define:}

\begin{equation}
(x',y',z')=S(x,y,z)
\label{eq:Reproduction}
\end{equation}
\noindent where $S$ is a function that takes as input a set of real world coordinates and outputs 3D projection coordinates.
Figure \ref{Reproduction} illustrates two examples.
Usually, specific algorithms \cite{40} are required to identify the data positioning based on the implicit physical structure of the data. Other algorithms may be employed to simplify intractable volumes and/or to derive additional features represented as colors, glyphs or streamlines. Reproduction can be seen as a special case of projection, where the projection function is unknown; instead, the data positioning derives from the observed phenomenon, as for volume rendering and geographical charts.\\

In reproduction, similarly to projections, data are mapped to positional marks that, via interpolation, can compose lines,
areas, surfaces or volumes - compare Figures \ref{Projection}(a) and
\ref{Reproduction}(a). Differently from projections, the positional reference (e.g., axes or background) is optional -- see Figure \ref{Reproduction}(a) -- and, usually, it is assumed that the graphical items are embedded in a Euclidean space.

From the expressiveness point of view, spatialization is the most important component of visualizations, providing both direct and derived perceptions. Direct perceptions include differentiation (natural for space, no positional mapping is required) and correspondence (a positional mapping is provided, as for projection and sequencing). Derived perceptions include arrangement (as for structure exposition) and/or meaning (as for projection and reproduction, which interpolate the points in space) occurring in parallel to other perceptions. 
Generally speaking, in visualization spatial design, the unique user decision, besides the spatialization method, is whether or not a positional reference will be used for correspondence.

\subsection{Shape}

\begin{figure*}[htb]
    \centering
\includegraphics[width=\ProportionOfText\textwidth]{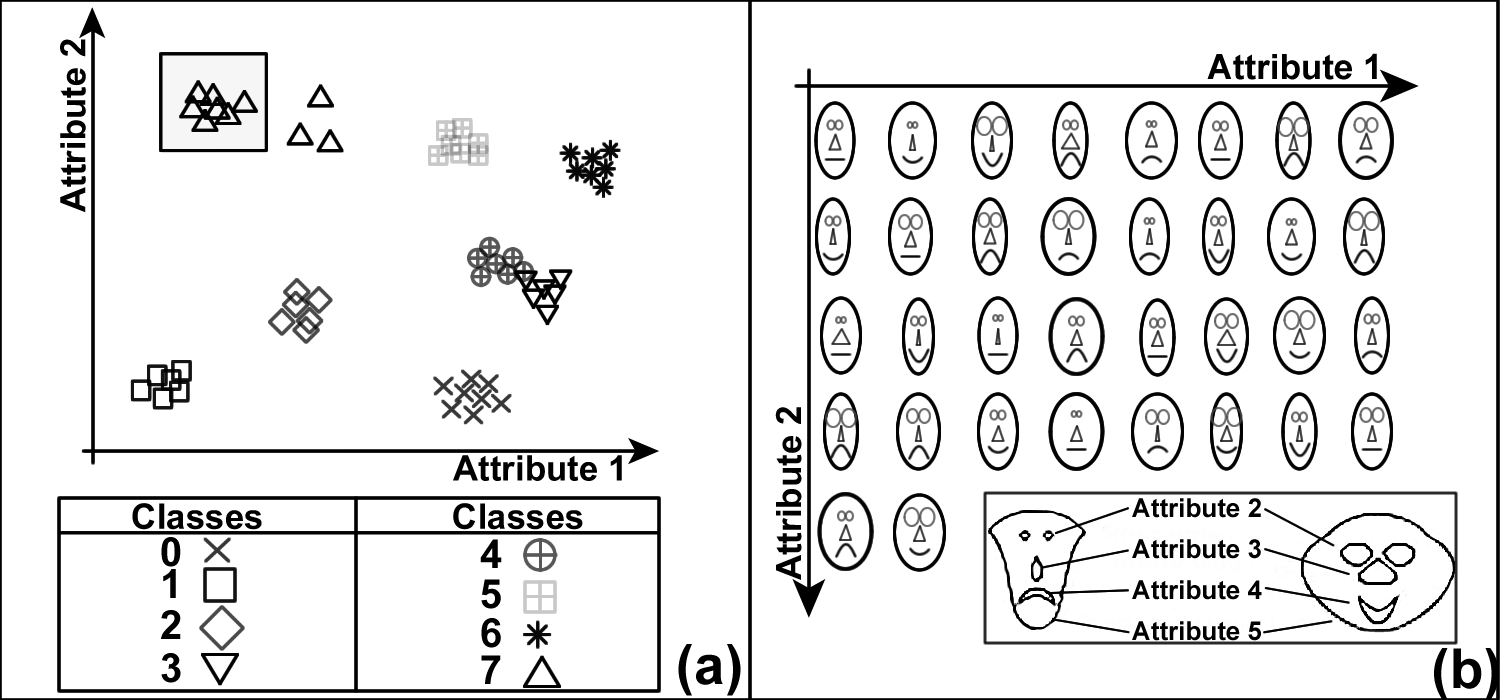}
    \caption{(a) Projection of classified items. Position: continuous correspondence to the projection axes; shape: discrete correspondence, and differentiation (selection square); color: none. (b) Chernoff faces projection. Position: continuous correspondence to the projection axes, and discrete correspondence to the features mapped to the human face; shape: continuous correspondence (size and curvature); color: none.}
    \label{Shape_ab}
\end{figure*}

\noindent{We have argued so far that a limited number of spatialization procedures is at the core of visualization techniques, and that these procedures dictate the positional pre-attentive stimulus.
Nevertheless, after spatializing the data, one still needs to decide how shape and color will compose the visualization. Hence, in this and in the following section we investigate how shape and color define visual perceptions for data interpretation -- as introduced in Section \ref{subsec:VisualExpressionProcess}. 
In particular, the Shape stimulus embraces the largest number of possibilities for visual perception: Correspondence, Differentiation, Connectivity and/or Meaning.}

\noindent{$\bullet${\it\ Correspondence}: discrete or continuous, each noticeable shape has a specific correspondence in a shape mapping. If no mapping is given, the intuitive mapping ``bigger size - higher magnitude'' is typically assumed.
	Figure \ref{Shape_ab} exemplifies the discrete (a) and the continuous cases (b);}   

\begin{figure*}[htb]
    \centering
\includegraphics[width=\ProportionOfText\textwidth]{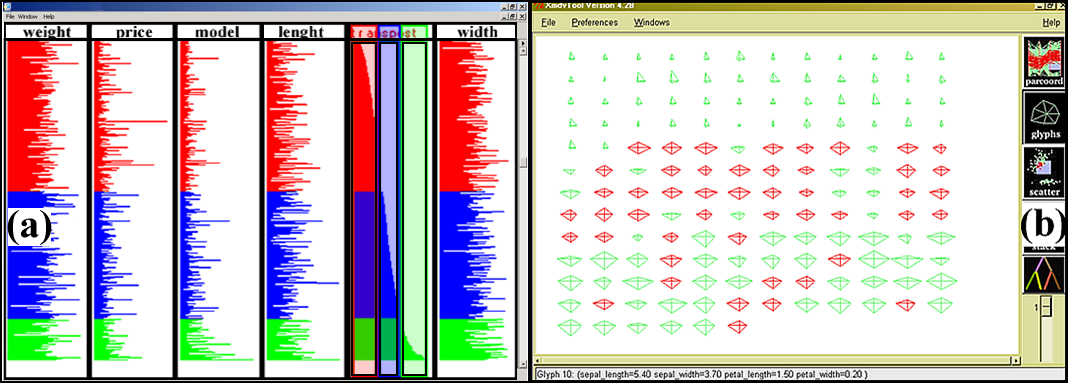}
    \caption{(a) Table Lens sequencing. Position: discrete correspondence to the axis labeled with attribute names in sequential positioning (horizontally), and discrete correspondence to the itens in sequential ordered positioning of the 5th column (vertically); shape: line width correspondence (all columns), and differentiation (on the 5th column, shapes indicate selection); color: discrete correspondence (colors correspond to selectors). (b) Star Glyphs projection. Position: correspondence to the window as an Euclidean projection plane, and correspondence through the circular sequencing of the inner sticks at each glyph; shape: differentiation determined by the contour around each glyph, and proportional correspondence for the inner sticks (length); color: discrete correspondence to selectors. (b) created with XmdvTool \cite{42}.}
    \label{Color_ab}
\end{figure*}

\noindent{$\bullet${\it\ Differentiation}: the displayed shapes simply discriminate the items for further interpretation, as in Figures \ref{Shape_ab}(a), \ref{Color_ab}(a) and \ref{ShapePosition}(a);}
    
\noindent{$\bullet${\it\ Connectivity}: line segments that denote connectivity between graphical items as, for example, in the Parallel Coordinates, in node-link graph visualization (illustrated in Figure \ref{StructureExposition}(b)), and in the visualization in Figure \ref{Position_ab}(a);}

\noindent{$\bullet${\it\ Meaning}: shapes such as arrows, faces or other complex formats (e.g. text) carry meaning whose interpretation relies on user knowledge, experience and culture, as depicted in Figures \ref{Reproduction}(b) and \ref{Position_ab}(b).}

At the border line between shape and position, shapes (arrows, crosses, triangles, faces, borders, etc.) are considered part of the visualization design only if they were explicitly selected to integrate it. Shapes not explicitly selected are in fact positional marks that may, via implicit or explicit interpolation, constitute lines, areas, surfaces or volumes, possibly with a certain complexity level that indirectly defines positional meaning.

\subsection{Color}

\noindent{After applying a spatialization procedure, which leads to positional cues to perceive information, and after choosing shape expressivity to
convey additional meaning, color is the third pre-attentive stimulus to be considered. Color conveys information through visual perceptions of Correspondence, Differentiation and/or Meaning:}

\begin{figure*}[htb]
    \centering
\includegraphics[width=\ProportionOfText\textwidth]{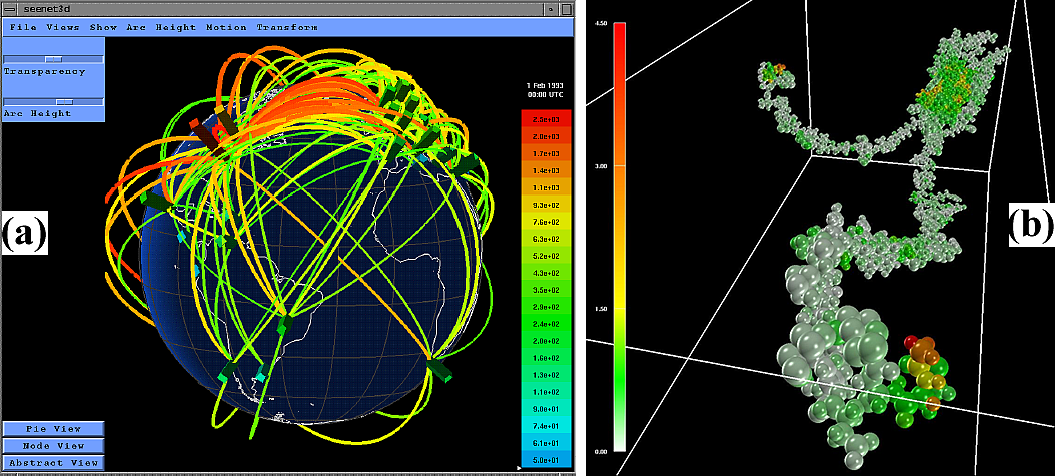}
    \caption{(a) Globe-map-based reproduction. Position: the reproduced points state continuous correspondence to Earthӳ globe; shape: connectivity (curved lines), and correspondence to pillar size; color: discrete correspondence. (b) Chemical structure reproduction. Position: the reproduced points state correspondence to the enclosing parallelepiped, and meaning according to the chemical structure achieved via interpolation of the points; shape: not given/chosen (explicit interpolation); color: continuous correspondence. Images reproduced with permission granted by Stephen G. Eick.}
    \label{Position_ab}
\end{figure*}

\noindent{$\bullet${\it\ Correspondence}: discrete or continuous. In the discrete case, each noticeable color defines a correspondence to a color mapping (see Figure \ref{Position_ab}(a)). In the continuous case, the variation of tones maps to a continuous range of values, as shown in Figure \ref{Position_ab}(b).}
    
\noindent{$\bullet${\it\ Differentiation}: colors bear no specific correspondence, they just depict an idea of equality (or inequality) of graphical entities, as it may be observed in the examples shown in Figures \ref{Sequencing}(b) and \ref{Reproduction}(b);}

\noindent{$\bullet${\it\ Meaning}: the displayed colors carry meaning. Examples include specific colors, e.g. red for alert, and specific materials resemblance (as in some textured colorings). The comprehension depends on the knowledge, experience and culture of the user.}

The Spatial-Perceptual Taxonomy is based on three criteria for classifying visualizations: space, shape and color. The set of these features defines different refinement perspectives for classification: only spatialization, only shape, only color and the combinations of these aspects. In the next sections, considering the possibility of multiple cycles of spatializations, we deepen into the notions of our taxonomy by defining its space of possibilities (the Spatial-Perceptual Design Space) and by defining a model for navigating this space (the Visualization Machine).

\subsection{The Taxonomy Formation Hierarchy}

\noindent{The proposed taxonomy classifies visualization techniques considering criteria that derive from the very constitution of the techniques. Hence, such criteria (spatialization, shape and color) are constituted by classes that reflect formation aspects. In our classification, a given technique falls within our taxonomy as an ensemble of factors that personify a constitutional categorization.}

\begin{figure*}[htb]
    \centering
\includegraphics[width=\ProportionOfText\textwidth]{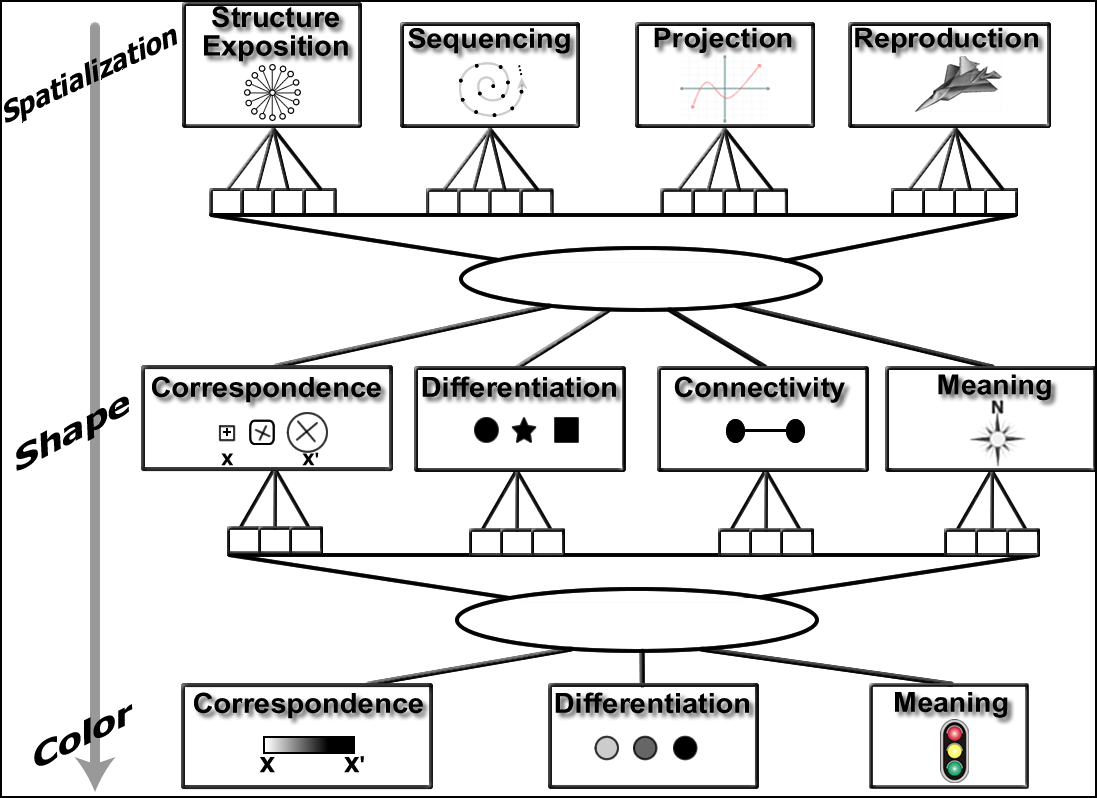}
    \caption{The formation hierarchy for the Spatial-Perceptual Taxonomy.}
    \label{fig:formation_tree}
\end{figure*}

Due to its descriptive -- rather than only classificatory -- properties and due to its number of criteria, the Spatial-Perceptual Taxonomy comes to a set of possibilities whose cardinality does not permit a readable enumeration of its classes, either in a tree or in a tabular format. For this reason, in figure \ref{fig:formation_tree}, we present the formation hierarchy for our taxonomy. This formation scheme permits to envision the classes that constitute our classification scheme via a top-down composition. This composition corresponds to the criteria and to the classes that compose our taxonomical organization.

\section{Analytical examples and multiple spatializations}
\label{sec:Analytical_examples}

\subsection{Analytical examples}
\label{Analytical_examples}

\begin{figure*}[htb]
    \centering
\includegraphics[width=0.48\textwidth]{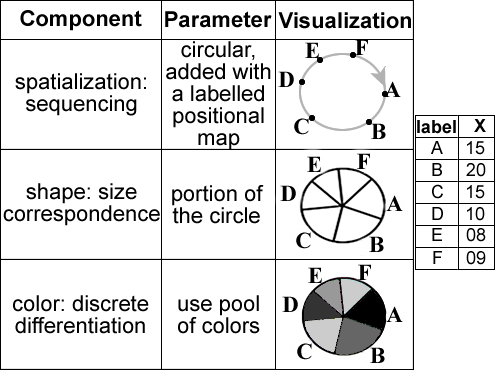}
    \caption{Components of a Pie Chart visualization.}
    \label{fig:Analysis1}
\end{figure*}

\noindent{Following, we illustrate our taxonomy by inspecting the components of two typical visualization techniques. 
First, let us consider a classical pie chart -- see Figure \ref{fig:Analysis1} -- along with a fictitious dataset. 
For this example, the visualization design starts by adopting a circular sequential spatialization along with a positional mapping of meaningful shapes (labels). 
The labels are used to state correspondences for each chart position. 
In a second step, shape (area) correspondence is applied to provide extra encoding. 
The result exhibits slices whose sizes are proportional to the values of attribute $X$. 
As the last step, a discrete coloring is applied for further differentiation.
The pie chart is an intuitive and straightforward visualization design.
For this reason, one tends to see it as a monolithic visualization, missing its components. 
Nevertheless, identifying such components may provide clues for new designs or to exploit design variations.}

\begin{figure*}[htb]
    \centering
\includegraphics[width=0.47\textwidth]{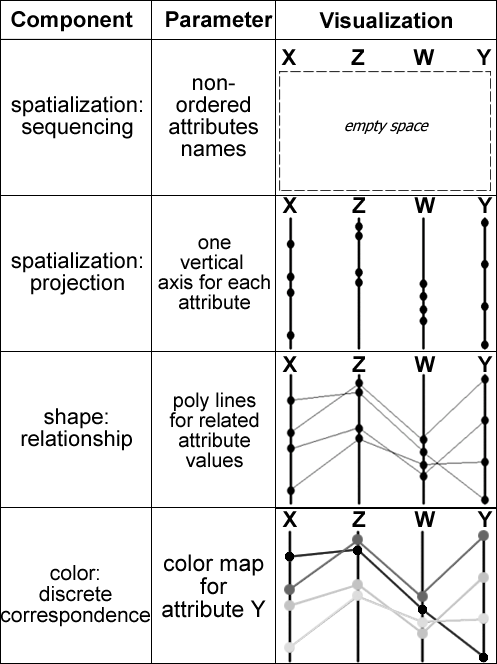}
    \caption{Components of a Parallel Coordinates visualization.}
    \label{fig:Analysis3}
\end{figure*}

In a second example, the Parallel Coordinates technique is used to illustrate the application of multiple spatialization procedures. 
We consider a hypothetical dataset with four-attribute ($W$, $X$, $Y$ and $Z$) records, where attribute $Y$ is a classification attribute. 
To start, a sequential positioning is defined for the names of the attributes. The low dimensionality enables having a positional map where the names of the attributes go along with the initial visualization. Each vertical position corresponds to a specific attribute name. 
In the second step, the values of each attribute are projected vertically having axes as positional maps. This second spatialization cycle benefits from the empty space in the display area. Note that the spatializations are integrated so that the items placed in the first cycle (attribute names) become a positional reference to the items placed in the second cycle (projected attribute values). In the next step of the design, shapes (line segments) stating connectivity are employed in order to express which data items are interrelated according to the dataset records. The final step uses color encoding in order to differentiate the polylines.

\subsection{Multiple spatializations}
\label{Hybridism}

\begin{figure*}[htb]
    \centering
\includegraphics[width=0.8\textwidth]{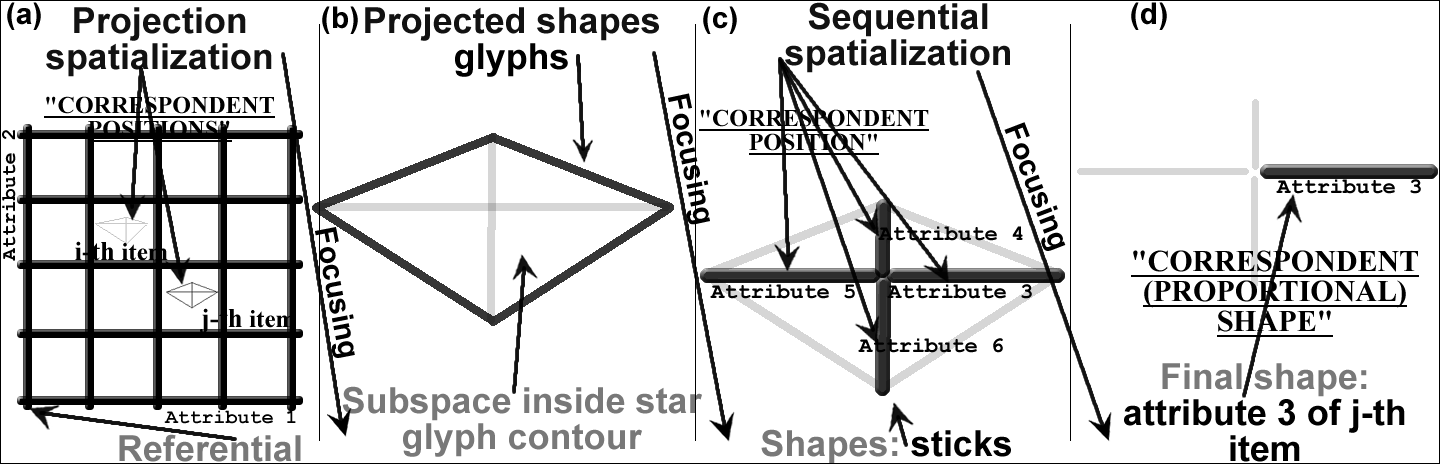}
    \caption{Two spatialization cycles applied in generating a visualization.
    (a) Projection of star glyphs. (b) Focus on a particular Star Glyph. (c) Sequencing within the glyph.
    (d) Attribute information as a shape correspondence.}
    \label{ShapePosition}
\end{figure*}

\noindent As exemplified in Figure \ref{fig:Analysis3}, visualizations may show disjoint regions, each with a different spatialization strategy. Another example can be seen in Figure \ref{ShapePosition}(a) for example, where it shows a grid in which star glyphs are spatialized according to a projection of two attributes. 
Figure \ref{ShapePosition}(b) focuses on one of these glyphs showing the available space within the glyph.
Figure \ref{ShapePosition}(c) shows that, within each glyph, the remaining attributes are positioned based on a Sequential spatialization. 
Finally, Figure \ref{ShapePosition}(d) focuses on a particular stick whose size corresponds to the magnitude of the third attribute of the (hypothetical) $j$-th item.

Similarly to the examples shown in Figures \ref{fig:Analysis3} and \ref{ShapePosition}, multiple spatialization cycles are applied in techniques such as Dimensional Stacking \cite{18}, Worlds-within-Worlds \cite{7}, Circle Segments
\cite{1}, Pixel Bar Charts \cite{16} and in many of the so-called iconic techniques. 
Multiple spatialization cycles define hybrid approaches that comprise a vast number of the techniques found in the visualization literature. 
In such compositions, pre-attention depends on how one focuses on the visualization.\\

Multiple spatialization cycles is a key factor for the diversity in visualization design. Integrated spatialization cycles allow improved space utilization and result in more complex techniques. In the next sections, we show that such understanding, coupled with our taxonomical system, can provide guidances on new thoughts for visualization. Table \ref{ClassificationExamplesTable} presents similar analysis for several visualization techniques widely referenced in the literature.

The case studies observed in table \ref{ClassificationExamplesTable} illustrate how the Spatial-Perceptual Taxonomy can analyze the constitution of several visualization techniques. 
Notably, one can perceive that the use of multiple (and possibly heterogeneous) spatializations are common, not to say necessary, in the visualization design. 
It is also possible to see that shape is not always explored, perhaps because of spatial limits. 
Color, in turn, can always be used without overloading the design framework. 
Each set of configurations in table \ref{ClassificationExamplesTable} can be understood as a complete characterization according to our taxonomy. 
Moreover, as pointed in section \ref{sec:Our_Taxonomy_Proposal}, the characterization that follows from our taxonomy can also be seen partially, considering for example, only classes based on spatialization or, for example, classes that arise from the combinations of shape and/or color.

\section{The Spatial-Perceptual Design Space}
\label{sec:DesignSpace}

\begin{figure*}[htb]
    \centering
\includegraphics[width=0.90\textwidth]{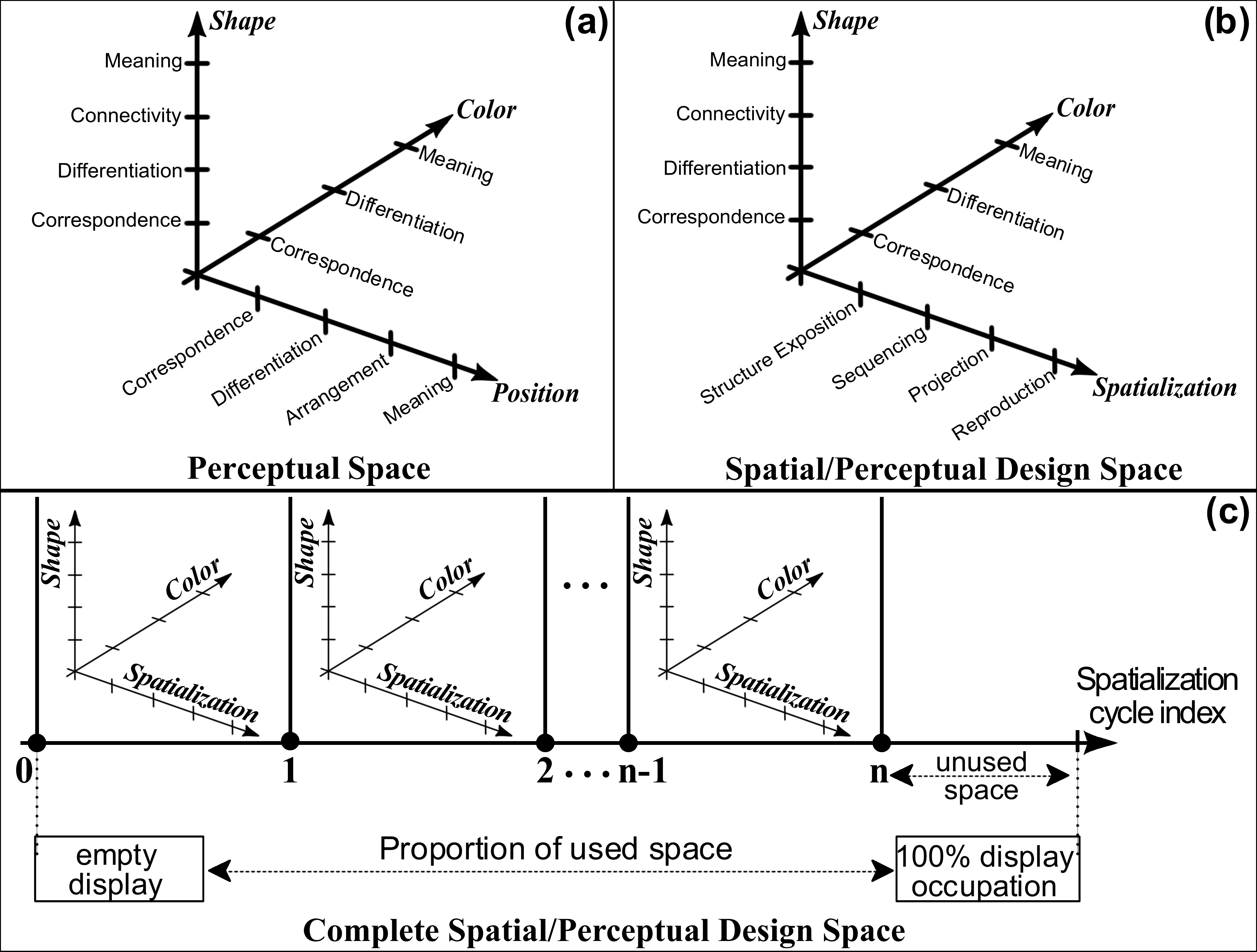}
    \caption{Design spaces. (a) The Perceptual Space with dimensions position, shape and color. The domains are subsets of the visual perceptions that we have identified. (b) The Spatial-Perceptual Design Space, where Position is dictated by the spatialization processes. (c) The complete Spatial-Perceptual Design Space, which includes multiple spatialization cycles.
}
    \label{fig:design_space}
\end{figure*}

\noindent{Based on the {\it Visual Expression Process} -- section \ref{subsec:VisualExpressionProcess} -- it is possible to conceive a {\it Perceptual Space}, as illustrated in Figure \ref{fig:design_space}(a), that describes the expressiveness of visualization techniques. The axes of this space correspond to the available possibilities of the basic visualization elements. As so, a visualization technique is defined in terms of its parameters regarding the choice of position, shape and color, each one cast to a subset of the visual perceptions introduced in Section \ref{subsec:VisualExpressionProcess}: correspondence, differentiation, connectivity, arrangement and meaning.}

This idea may be further refined under the {\it Spatial-Perceptual Taxonomy} introduced in Section \ref{sec:Our_Taxonomy_Proposal}, which supports the definition of a design space for the visualization techniques. 
The {\it Spatial-Perceptual Design Space}, illustrated in Figure \ref{fig:design_space}(b), assumes that positional perception is dictated by the spatialization processes, as structured in our taxonomy. In this design space, the positional dimension becomes the spatialization dimension, according to how we describe this concept in Section \ref{sec:Our_Taxonomy_Proposal}.

Finally, in a third step, we consider the possibility of multiple spatialization cycles, as discussed in Section \ref{Hybridism}, to depict the complete spatial-perceptual design space. The design process now becomes a sequence of space filling cycles that follows the space/perception design and proceeds until all the available space is occupied. In Figure \ref{fig:design_space}(c) we present the Spatial-Perceptual Design Space for techniques whose conception adopts multiple spatialization cycles.

While the Spatial-Perceptual Taxonomy identifies the categories of visualizations features, the Spatial-Perceptual Design Space presents this notion in an intuitive Euclidean space representation, where each point addresses a possible class. In this line, our motivation is to make clear what can be done in terms of the visualization design and how the corresponding possibilities can be organized from a perceptual-cognitive perspective. The overall idea follows two steps:

\begin{enumerate}
	\item{Make choices for spatialization, shape and color in terms of correspondence, differentiation, connectivity, arrangement and meaning;}
	\item{Iterate through multiple cycles while there is available space.}
\end{enumerate}

In the next sections we also observe that, for complete visualization systems (data management + visualization + interaction), other two steps are necessary: pre-exhibition data processing and interaction.

\section{Interaction Techniques}
\label{sec:InteractionTechniques}

\noindent{The scope that we have chosen to explore in this work is {\em the definition of a design space oriented to visual expressivity}. Accordingly, we have focused on the visual appeal of sole data visualization techniques, rather than on the operational features of a complete visualization system (data management + visualization + interaction).}

Unlike previous works, interaction is not a component of our theory, rather, it rises as a natural product of the proposed concepts. In this section, we clarify the role of interaction techniques under the light of the proposed ideas. However, this work does not intend to determine {\em how} interaction can be used in the visualization design: our aim is only to determine {\em where} does interaction fit our theory. Interaction is a much wider field of research, thus we believe that separating interaction from visualization helps to keep the conceptual model clearer. Restricting the analysis to the domain of techniques for data visualization, we establish two conditions to identify an interaction technique:

\begin{enumerate}
    \item{An interaction technique must enable a user to define/redefine the visualization by modifying pre-attentive stimuli;}
    \item{An interaction technique, with appropriate adaptations, must be applicable to any visualization technique, in an efficient way or not.}
\end{enumerate}

The first condition is a direct consequence of the assumption that interaction techniques changes the state of a computational application. In the case of a visualization scene, its basic components (the pre-attentive stimuli) must be altered. 
The second condition derives from the need to have a well-defined, yet general, concept. Therefore, interaction techniques, then, must be applicable to any visualization technique, even if not efficiently. In the current literature, we identify the following interaction paradigms satisfying our criteria:

\begin{compactitem}
    \item{\it Parametric}: the visualization is indirectly redefined through mechanisms that reflect on new parameters for position, shape or color; visually (e.g., scrollbar) or textually (e.g., type-in). An example is the Hierarchical Brushing mechanism described by Fua {\it et al.} \cite{8};
    \item{{\it View transformation}: this interaction allows changing shape (size) and position of a visual scene through scale, rotation, translation and/or zoom, as in FastmapDB \cite{33};}
    \item{{\it Filtering}: a user can visually select a subset of items that will be promptly differentiated for user perception by changing pre-attentive properties such as color (brushing) and shape (selection contour). Detailed studies are presented by Martin and Ward \cite{20};}
    \item{{\it Details-on-demand}: detailed information about the data that generated a particular visual entity can be retrieved for exhibition. As an example, we refer to the interaction used in the Table Lens visualization technique -- it allows retrieving the data that originated a given graphical item and present it using textual (shape) visual patterns;}
    \item{{\it Distortion}: allows visualizations to be projected so that different perspectives (positions) can be observed and defined simultaneously. Classical examples are the Perspective Wall \cite{19} and the Fish-eye Views \cite{28}.\\}

\end{compactitem}

The well-known Link \& Brush (co-plots) technique does not satisfy the conditions to be considered an interaction technique. Link \& Brush is more like a design dependent automation. It is based on the possibility of integrating multiple spatializations so that interaction tasks applied to one of the spatializations reflect on the others. 
One could generalize Link \& Brush and think of Link \& View Transformation (e.g, synchronized panning), Link \& Parameters (e.g, parameters multicasting), Link \& Distort (e.g., simultaneous perspective alteration), and so forth.

\section{The Visualization Machine model}
\label{sec:GeneralVisualization}

\begin{figure*}[htb]
    \centering
\includegraphics[width=\textwidth]{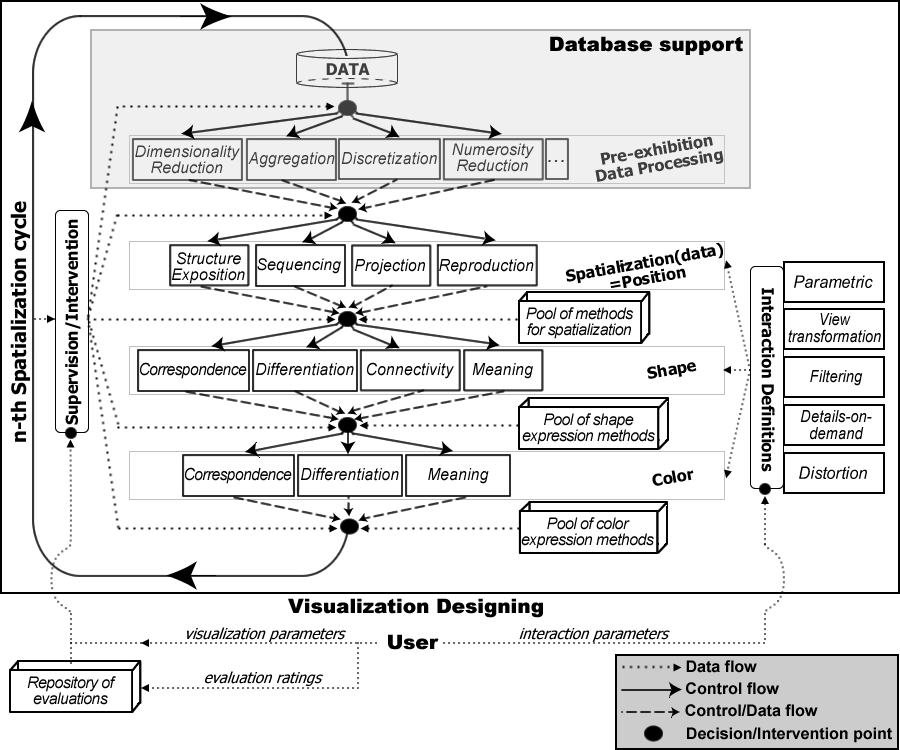}
    \caption{The Visualization Machine model for general visualization design.}
    \label{fig:GeneralVisualization}
\end{figure*}

\noindent{Facing visualizations from the perspective of the proposed theory allows conceiving an ideal generalized
visualization mechanism. This is illustrated in Figure \ref{fig:GeneralVisualization}, which shows a model integrating the concepts described so far.\\}

The Visualization Machine, described next, is not a finalized scheme, but a seminal idea, a new way of thinking. 
Its principle is the reduction of the visualization design in such a way that, theoretically, it could be made mechanically.
Its description personifies both the Spatial-Perceptual Taxonomy and the Spatial-Perceptual Design Space. The framework delineates a model for navigating through the possibilities of design predicted by our theory. In this context, the frame over which the Visualization Machine evolves is in the format of a hierarchy of possibilities. The branches of these hierarchy lead to classifications based on spatialization, shape and color.

\noindent{\bf A top-down model\\} Our model states that a visualization is achieved by carrying out a sequence of steps that starts with the spatialization of a set of data items. 
The initial spatialization then undergoes shape and color encodings and may be followed by other spatialization cycles. 
In the model illustrated in Figure \ref{fig:GeneralVisualization}, each step of the visualization design defines a decision point (black circles in the figure) that will iteratively generate a visualization that conforms to the choices of the user. 
A careful look reveals that the Visualization Machine is a model that supports user navigation through the space defined by the Spatial-Perceptual Design Space presented in section \ref{sec:DesignSpace}.\\

A pre-exhibition data processing step complements the overall design process. The pre-exhibition data processing is not in the main scope of this work, which is the definition of a design space oriented to visual expressivity. However, as the pre-exhibition data processing is needed for many visualization systems, we have added it to our conceptual model. This processing step occurs before the spatialization, when data are usually processed to enable better management and mapping. The step is executed using database support and includes operations such as the reduction of the number of data items, selection of attributes, dimensionality reduction, computing summaries, classify data items, perform data mining operations, to name a few. In many domains, pre-exhibition data processing is a necessary step before conceiving the visualization.\\

\noindent{\bf Pools of parameters\\} In our systematization, the steps for spatialization, shape and color encodings are assisted, respectively, by pools of spatialization methods, pools of shape expression methods and pools of color expression methods. The pools represent sets of known methods that can be used to conceive patterns for visual ensembles. User intervention permits to use these pools to define parameters for each of the steps, from pre-exhibition data processing to coloring, allowing the user to choose the procedures and the patterns that best match her/his needs. Along the text we have shown several examples of design that apply to well-known visualization techniques. The constituents of such designs exemplify the repertory of the pools of parameters that we propose.\\

\noindent{\bf Supervision/Intervention\\} The choice of parameters defined via user intervention demands supervision, both automatic and user provided. 
A supervision module is depicted at the left-hand side of the schema in Figure \ref{fig:GeneralVisualization}. 
Supervision is aimed at validating and verifying the parameters for each step of the design process in order to determine what are the choices that can be applied in the current and in the subsequent steps. 
In the current step, supervision is supposed to consider the current user choice in order to select the methods that can be properly used from the current pool of parameters. For the subsequent steps, supervision is supposed to filter out choices that might lead to bad designs. 
For example, due to the limited space in conventional displays, it is not reasonable to choose a sequential spatialization of a million items from a dataset and then to choose a shape correspondence in the next step. 
More important, the supervision module is also cast to collect data about the spatialization cycles. 
This is necessary in order to monitor the available space to be used by further spatialization cycles, and to detect and offer possibilities to integrate multiple spatializations. Such integration is essential for the conception of more complex techniques that benefit from multiple spatializations. For example, it can promote link-like automation, so that interaction parameters are multi-cast to more than one spatialization; or it can promote coordinated spatializations, so that the items in different display areas follow the same ordering.\\

\hyphenpenalty=10000
\begin{table*}[htb]
  \begin{center}
  \caption{Examples of spatial-perceptual analyses considering common settings for classical techniques.}
  \label{ClassificationExamplesTable}
  \begin{tabular}{|p{1.51in}|p{1.61in}|p{1.07in}|p{0.75in}|p{1.06in}|}
    \hline    	 
          {\footnotesize{\it{Visualization Technique}}}
         &{\footnotesize{\it{Spatialization}}}
         &{\footnotesize{\it{Shape}}}
         &{\footnotesize{\it{Color}}}
         &{\footnotesize{\it{Prospective Interaction}}}\\
    \hline

    \hline    
      {\footnotesize{Chernoff Faces \cite{4}}}
     &{\footnotesize{Projection, Sequencing}}
     &{\footnotesize{Differentiation, Correspondence}}
     &{\footnotesize{-}}
     &{\footnotesize{Filtering}}\\
     
    \hline 
      {\footnotesize{\mbox{Dimensional Stacking} \cite{18}}}
     &{\footnotesize{\mbox{Multiple Projection}}}
     &{\footnotesize{-}}
     &{\footnotesize{Differentiation}}
     &{\footnotesize{Filtering}}\\
     
    \hline 
      {\footnotesize{Parallel Coordinates \cite{10}}}
     &{\footnotesize{Sequencing, \mbox{Multiple Projection}}}
     &{\footnotesize{Connectivity}}
     &{\footnotesize{Differentiation}}
     &{\footnotesize{Filtering}}\\
     
    \hline
      {\footnotesize{Scatter Plots \cite{6}}}
     &{\footnotesize{\mbox{Multiple Projection}}}
     &{\footnotesize{-}}
     &{\footnotesize{Differentiation}}
     &{\footnotesize{Filtering}}\\
     
    \hline
      {\footnotesize{Star Coordinates \cite{12}}}
     &{\footnotesize{Projection}}
     &{\footnotesize{-}}
     &{\footnotesize{Differentiation}}
     &{\footnotesize{Filtering, \mbox{View transformation}}}\\
     
    \hline
       {\footnotesize{Stick Figures \cite{22}}}
      &{\footnotesize{Projection, \mbox{Multiple Sequencing}}}
      &{\footnotesize{Differentiation, Correspondence}}
      &{\footnotesize{Differentiation}}
      &{\footnotesize{Filtering}}\\
      
    \hline
       {\footnotesize{Worlds-within-Worlds \cite{7}}}
      &{\footnotesize{\mbox{Multiple Projection}}}
      &{\footnotesize{-}}
      &{\footnotesize{Differentiation}}
      &{\footnotesize{\mbox{View transformation}}}\\
      
    \hline      
      {\footnotesize{Parallel Coordinates/Star Glyphs \cite{FaneaGlyphCoordinates05}}}
     &{\footnotesize{\mbox{Multiple Sequencing}, \mbox{Multiple Projection}}}
     &{\footnotesize{Connectivity}}
     &{\footnotesize{Correspondence}}
     &{\footnotesize{Filtering, parametric}}\\
     
    \hline
      {\footnotesize{{\it Flow Map Layout} \cite{PhanFlowMap05}}}
     &{\footnotesize{Projection}}
     &{\footnotesize{Correspondence, Connectivity}}
     &{\footnotesize{Correspondence}}  &{\footnotesize{\mbox{Details-on-demand}}}\\
     
    \hline      
      {\footnotesize{Bar Chart}}
     &{\footnotesize{Projection}}
     &{\footnotesize{Correspondence}}
     &{\footnotesize{Correspondence}}
     &{\footnotesize{Filtering, Parametric}}\\
     
    \hline
      {\footnotesize{Pixel Bar Charts \cite{16}}}
     &{\footnotesize{Projection, \mbox{Multiple Sequencing}}}
     &{\footnotesize{Correspondence}}
     &{\footnotesize{Correspondence}}
     &{\footnotesize{Filtering, Parametric}}\\
     
    \hline
      {\footnotesize{Circle Segments \cite{1}}}
     &{\footnotesize{\mbox{Multiple Sequencing}}}
     &{\footnotesize{Differentiation}}
     &{\footnotesize{Correspondence}}
     &{\footnotesize{Filtering, \mbox{Details-on-demand}}}\\
     
    \hline
      {\footnotesize{Pixel Oriented Techniques by Keim \cite{13}}}
     &{\footnotesize{\mbox{Multiple Sequencing}}}
     &{\footnotesize{-}}
     &{\footnotesize{Correspondence}}
     &{\footnotesize{Filtering, Parametric, \mbox{Details-on-demand}}}\\
     
    \hline
      {\footnotesize{Pie Chart}}
		 &{\footnotesize{Sequencing}}
		 &{\footnotesize{Correspondence}}
		 &{\footnotesize{Differentiation}}
		 &{\footnotesize{Filtering, Parametric}}\\
		 
    \hline
      {\footnotesize{Table Lens \cite{24}}}
     &{\footnotesize{Sequencing, \mbox{Multiple Sequencing}}}
     &{\footnotesize{Correspondence}}
     &{\footnotesize{Differentiation}}
     &{\footnotesize{Filtering, \mbox{Details-on-demand}}}\\
     
    \hline
      {\footnotesize{{\it InterRing} \cite{YangInterRing02}}}
     &{\footnotesize{Structure Exposition}}
     &{\footnotesize{Correspondence}}
     &{\footnotesize{Correspondence}}
     &{\footnotesize{\mbox{View transformation}, \mbox{Details-on-demand}}}\\
     
    \hline
      {\footnotesize{Cone Tree \cite{26}}}
     &{\footnotesize{Structure Exposition}}
     &{\footnotesize{Connectivity}}
     &{\footnotesize{Differentiation}}
     &{\footnotesize{\mbox{View transformation}, \mbox{Details-on-demand}}}\\
     
    \hline
      {\footnotesize{Hyperbolic Tree \cite{17}}}
     &{\footnotesize{Structure Exposition}}
     &{\footnotesize{Connectivity}}
     &{\footnotesize{Differentiation}}
     &{\footnotesize{\mbox{View transformation}, \mbox{Details-on-demand}}}\\
     
    \hline
      {\footnotesize{Treemaps \cite{30}}}
     &{\footnotesize{Structure Exposition}}
     &{\footnotesize{Correspondence}}
     &{\footnotesize{Differentiation}}
     &{\footnotesize{Filtering, \mbox{Details-on-demand}}}\\
     
    \hline
      {\footnotesize{Voronoi {\it Tree-maps} \cite{BalzerTreemaps05}}}
     &{\footnotesize{Structure Exposition}}
     &{\footnotesize{Correspondence}}
     &{\footnotesize{Correspondence}}
     &{\footnotesize{Filtering, \mbox{Details-on-demand}}}\\
     
    \hline     
      {\footnotesize{Geographical Maps}}
     &{\footnotesize{Reproduction with referential}}
     &{\footnotesize{Differentiation, Meaning}}
     &{\footnotesize{Differentiation, Correspondence}}
     &{\footnotesize{\mbox{View transformation}, \mbox{Details-on-demand}}}\\
     
    \hline
      {\footnotesize{Vector Visualization}}
     &{\footnotesize{Reproduction}}
     &{\footnotesize{Meaning, Correspondence}}
     &{\footnotesize{Correspondence}}
     &{\footnotesize{\mbox{View transformation}}}\\
     
    \hline
      {\footnotesize{\mbox{Direct Volume Rendering} \cite{36}}}
     &{\footnotesize{Reproduction}}
     &{\footnotesize{Not given/chosen \mbox{(explicit interpolation)}}}
     &{\footnotesize{Correspondence}}
     &{\footnotesize{\mbox{View transformation}}}\\
     
    \hline    
	\end{tabular}
  \end{center}	
\end{table*}	
 \hyphenpenalty=50

\noindent{\bf Interaction\\} The definition of interaction introduced in Section \ref{sec:InteractionTechniques} states that
any interaction technique fits any visualization. 
Thus, any interaction technique should be readily applicable in the Visualization Machine model. 
Of course, the availability and adaptation of each interaction paradigm will follow the nature of the specific parameters (position, shape and color) chosen for the design. Interaction efficiency reflects the integration of all the selected components.\\

\noindent{\bf Evaluation\\}
The parameter-based top-down Visualization Machine model ensures a high level of freedom in combining multiple encoding components. Such combinations can be efficient, reasonable or even catastrophic, depending on the goals of the user who defines the visualization setting. 
Therefore, the model also embodies a repository of evaluations (lower left in the diagram). Once a user defines a visualization setting and is able to use it, she/he can evaluate it according to her/his goals. Each evaluation is stored at the repository module that joins the information from the dataset being analyzed, from the interaction techniques and from the parameters that define the visualization setting. The information becomes available to the supervision module for user assistance and validation, and for automatic parameter definition during subsequent system usage.\\

\noindent{\bf An expanding self evaluation/learning system\\} 
The Visualization Machine model is structured as a system that can create visualizations based on sets of design parameters. 
It can be considered an environment that provides and manipulates aspects of the data visualization science. Under this perspective, visualization research may be seen as a discipline that tackles discovery and empirical evaluation of spatialization and expression methods that fit the proposed model. 
New methods would be incrementally incorporated and established in the model, at the repository pools. In this scenario, it is possible to think of an incremental visualization environment in which users would apply parameters previously conceived by specialist designers. This possibility contrasts with the current paradigm, in which visualizations are limited to implementations following specific, rigid designs.\\
The self evaluating characteristic of the model grants it to be a learning system that adapts to different user profiles or that converges towards a universal visualization tool.\\

\noindent{\bf A Conceptual Proposition}\\
The Visualization Machine is idealized as a tool to enable browsing collections of methods. It is envisioned as a composing process that aims at expressing visual perceptions of correspondence, differentiation, connectivity, arrangement and meaning using space, shape and color. At the same time that the user would be able to browse and combine collections of methods, the machine would monitor possible combinations of methods by suggesting previously successful configurations and preventing non-prospective ones. This course of conception is conceptual and plenary resulting from the consideration of our theory. Not surprisingly, its realization raises technical challenges. It is necessary a clear interface for universally accommodating methods (especially interaction methods) providing an expanding framework, each time more complete. It is also desirable a well-defined standard for recording the configuration sets of the machine and the evaluation of the users in terms of the data interpretation.

Together with the Spatial-Perceptual Taxonomy and Design Space, the Visualization Machine defines a set of concepts for an alternative understanding of the data visualization design. It is not a complete project for development but, rather, it is a ``conceptual proposition'' intended to propel new perspectives for visualization. Its principle is to consider visualizations as combinations of a repetitively employed and limited (though not small) set of methods and techniques categorized in terms of their nature (space, shape and color) and expressivity (visual perceptions). It is supposed to be a new framework where new implementations can be accommodated and combined with already existing modules. We believe this is the overall line through which visualization science will evolve, probably reaching a unified convergence.

Despite the idealized nature of the model, which hides many intricacies, we observe that its completeness and coherence provide a starting point to accomplish such a tool or to conceive a framework on a similar basis. Following we describe our design process through two examples.\\

\noindent{{\bf Star Glyphs}\\}

\begin{figure*}[htb]
    \centering
\includegraphics[width=0.6\textwidth]{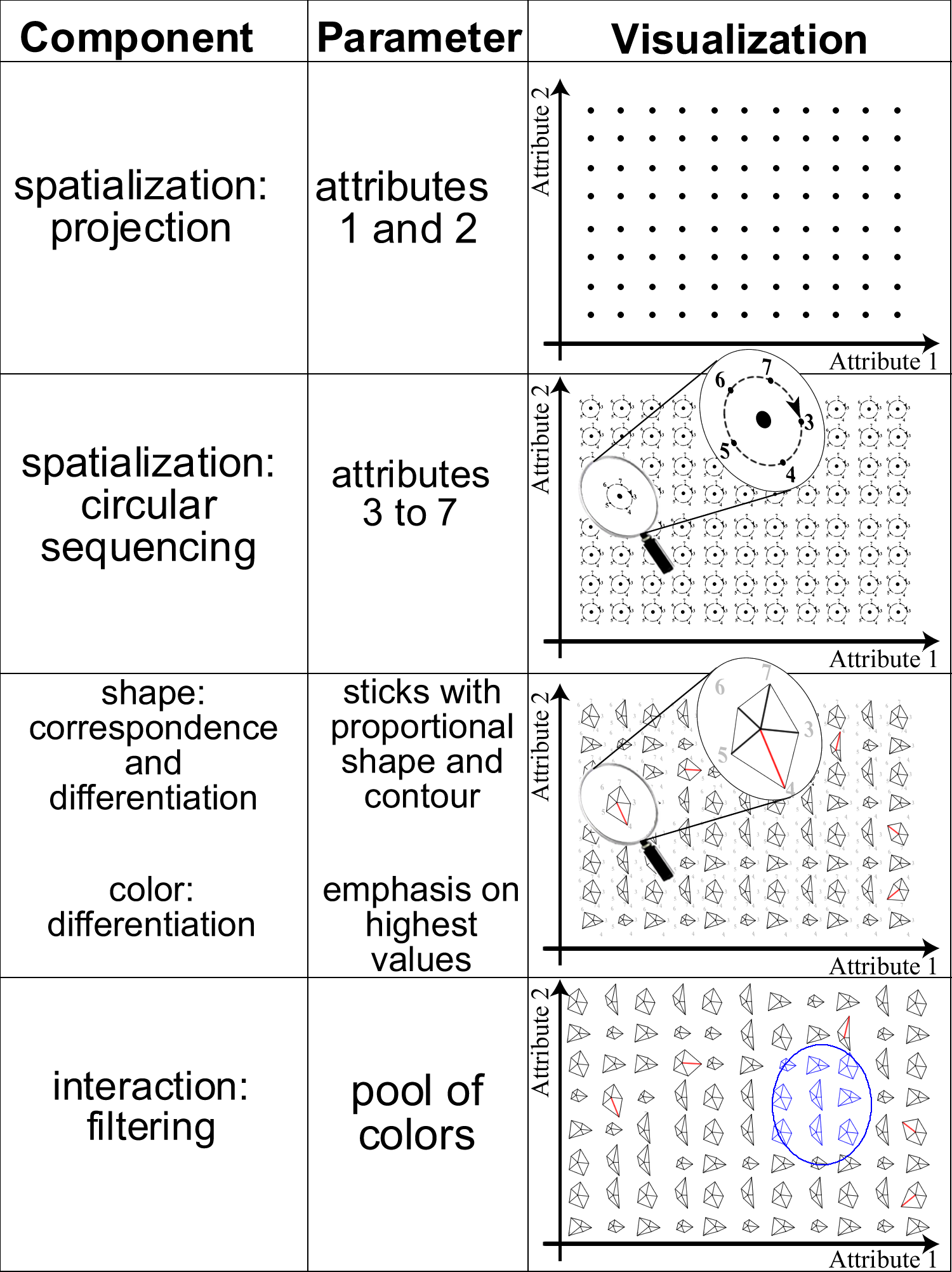}
    \caption{Design process for classic Star Glyphs technique according to the Visualization Machine model.}
    \label{fig:StarGlyphsDesign}
\end{figure*}

\noindent{As a first example, we go through the design process of the classic Star Glyphs technique, illustrated in Figure \ref{fig:StarGlyphsDesign}. The design conception starts by adopting a spatial projection according to attributes, say, 1 and 2; no shape; no color. These first decisions define a complete cycle over the Visualization Machine logic. Thereafter, a second cycle takes place. For each projected data item, and respecting the remaining empty space, we apply a sequential circular spatialization of attributes 3 to 7; for shape encoding, we opt for correspondence and choose sticks with proportional shape; we opt for differentiation connecting the outer extremities and determining the contour of the sticks; for color, we can benefit from differentiation, having the highest value in each dimension emphasized in red. Finally, having two cycles of design, we choose to add interaction to the first cycle adding filtering over the projected items.}\\

\noindent{{\bf Coordinated Parallel Bar Charts}}

\begin{figure*}[htb]
    \centering
\includegraphics[width=0.6\textwidth]{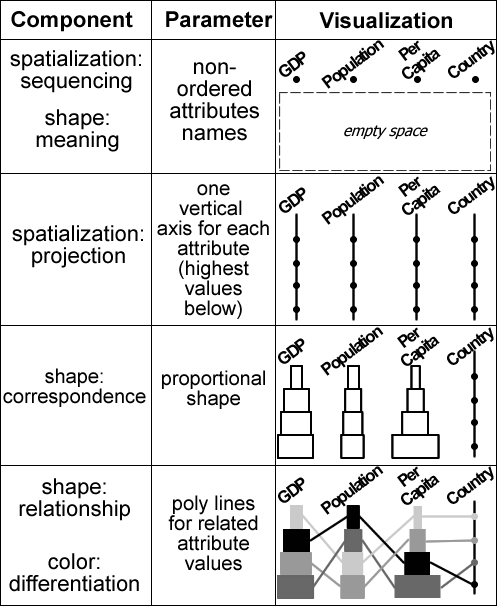}
    \caption{A design process derived from the Parallel Coordinates technique, named Coordinated Parallel Bar Charts.}
    \label{fig:ParallelBarCharts}
\end{figure*}

\noindent{In this example (see Figure \ref{fig:ParallelBarCharts}), we design a variation of the well-known Parallel Coordinates technique called Coordinated Parallel Bar Charts. Initially we consider a semantically-rich hypothetical dataset with four-attribute records (gross domestic product, population, per capita income and country name). First, like in the Parallel Coordinates technique, we apply a sequential positioning of the attributes; meaningful shapes (labels) and no color. In a second cycle, the values of each attribute are projected vertically, having the highest values at the lower part of the axes. At this point we observe that it is possible to benefit from existing correspondences using shape proportionality. This decision creates the appearance of a sequence of bar charts presentation. In the last steps, we choose relationship via shape encoding (polylines) and color differentiation much like in the classic Parallel Coordinates. The final result is a technique that, besides showing the values of the attributes through positional correspondence, also stresses the distortions that can be observed in the domain of each attribute.}

\section{Conclusions}
\label{sec:Conclusions}

\noindent{In this paper we have introduced a new comprehensive perspective for visualization design. To do so, initially we identified the Visual Expression Process in order to delineate a new strategy for analysis. This observation allowed us to concentrate on the expressiveness of the visualizations in order to identify their elements. The elements are the basis to conceive the Spatial-Perceptual Taxonomy, which considers visualizations as sets of components that afford a limited number of visual perceptions. 
This investigation led us to the Spatial-Perceptual Design Space, which aggregates the presented ideas in a single conception. For completeness, we have also discussed how interaction techniques fit into our theory. The proposed taxonomy and design space were finally integrated into a design scheme named Visualization Machine Model. Along the text, to validate our observations, we analytically reviewed a number of designs according to the proposed ideas.}

The proposed work is centered on how users perceive visualizations, rather than on which patterns are used to build a given visualization. Unlike former approaches that depart from the set of available visual patterns, we ground our ideas on the set of visual perceptions, in order to choose approaches for position, shape and color encoding. Our approach brings designers closer to the expressive characteristics of visual ensembles. We believe that this theorization can promote a more intuitive understanding of the design process, fostering further research toward a more precise and comprehensive design science concerning visualization techniques.

This work introduces straight future research lines. First, the Visualization Machine is a conceptual proposition to illustrate the usability of our theory. Its feasibility and complexity must be traced and compared to other visualization libraries and packages. Another line of research concerns the topic of animation, which applies to all the features studied in this work: space, shape and color. The study of animation requires an ample understanding of its implications and of the intricacies of its elaboration. Finally, a similar study must be carried out at the level of visualization systems so that, expanding our line of analysis beyond visual design, the structuring and construction of complex visualization environments can be envisioned with animation, advanced interaction and multiple coordinated views.

\bibliographystyle{plain}
  {\small
    {

    }
}

\end{document}